\documentclass{aa}
\usepackage{graphicx}
\usepackage[varg]{txfonts}

\usepackage{natbib}
\usepackage[breaklinks=true, colorlinks=true, linkcolor=blue, citecolor=blue]{hyperref}

\usepackage{multirow}

\begin{document}

\title{X-ray studies of the Abell 3158 galaxy cluster with eROSITA}

\titlerunning{A3158}

\author{B. Whelan \inst{1} \and A. Veronica\inst{1} \and F. Pacaud \inst{1} \and T. H. Reiprich \inst{1}  \and E. Bulbul \inst{2} \and M. E. Ramos-Ceja\inst{2} \and J. S. Sanders\inst{2} \and J. Aschersleben\inst{1} \and J. Iljenkarevic\inst{1} \and K. Migkas\inst{1} \and M. Freyberg\inst{2} \and K. Dennerl\inst{2} \and M. Kara\inst{1,4} \and A. Liu\inst{2} \and V. Ghirardini\inst{2} \and N. Ota\inst{1,3} } 

\institute{Argelander-Institut f\"ur Astronomie (AIfA), Universit\"at Bonn,
              Auf dem H\"ugel 71, 53121 Bonn, Germany\\
              \email{reiprich@astro.uni-bonn.de}
         \and
             Max-Planck-Institut f\"ur extraterrestrische Physik, Giessenbachstra{\ss}e 1, 85748 Garching, Germany
             \and 
            Department of Physics, Nara Women's University, Kitauoyanishi-machi, Nara, 630-8506, Japan
            \and
            Institute for Astroparticle Physics, Karlsruhe Institute of Technology, 76021 Karlsruhe, Germany}
\date{Received \dots; Accepted \dots}

\abstract
{The most nearby clusters are the best places for studying physical and enrichment effects in the faint cluster outskirts. The Abell 3158 cluster (A3158), located at $z=0.059,$ is quite extended with a characteristic radius $r_{200} = 23.95$\,arcmin.  The metal distribution in the outskirts of this cluster has previously been studied with \textit{XMM-Newton}. In 2019, A3158 was observed as a calibration target in a pointed observation with the eROSITA telescope on board the Spektrum-Roentgen-Gamma mission. Bright large clusters, such as A3158, are ideal for studying the metal distribution in the cluster outskirts, along with the temperature profile and morphology. With the deeper observation time of the eROSITA telescope, these properties can now be studied in greater detail and at larger radii. Furthermore, bright nearby clusters are ideal X-ray instrumental cross-calibration targets as they cover a large fraction of the detector and do not vary in time.}
{We first compare the temperature, metal abundance, and normalisation profiles of the cluster from eROSITA with previous \textit{XMM-Newton} and \textit{Chandra} data. Following this calibration work, we investigate the temperature and metallicity of the cluster out to almost $r_{200}$, measure the galaxy velocity dispersion, and determine the cluster mass. Furthermore, we search for infalling clumps and background clusters in the field.}
{We determined 1D temperature, abundance,
and normalisation profiles from both eROSITA and \textit{XMM-Newton} data as well as 2D maps of temperature and metal abundance distribution from eROSITA data. The velocity dispersion was determined and the cluster mass was calculated from the mass - velocity dispersion ($M_{200}$ - $\sigma_{v}$) relation. Galaxy density maps were created to enable a better understanding of the structure of the cluster and the outskirts.} 
{The overall (i.e. in the range $0.2 - 0.5 r_{500}$) temperature was measured to be $5.158 \pm 0.038$\,keV. The temperature, abundance, and normalisation profiles of eROSITA all agree to within a confidence level of about 10\% with those we determined using \textit{XMM-Newton} and \textit{Chandra} data, and they are also consistent with the profiles published previously by the X-COP project. The cluster morphology and surface brightness profile of cluster Abell 3158 appear to be regular at a first glance. Clusters that have such profiles typically are relaxed and host cool cores. However, the temperature profile and map show that the cluster lacks a cool core, as was noted before. Instead, an off-centre cool clump lies to the west of the central cluster region, as reported previously. These are indications that the cluster may be undergoing some sloshing and merger activity. Furthermore, there is a bow-shaped edge near the location of the cool gas clump west of the cluster centre. Farther out west of the X-ray images of A3158, an extension of gas is detected. This larger-scale extension is described here for the first time.
The gas metallicity ($\sim$0.2 solar) measured in the outskirts ($\gg$$r_{500})$ is consistent with an early-enrichment scenario.
The velocity dispersion of the cluster member galaxies is measured to be $1058 \pm 41 {\rm km\,s}^{-1}$ based on spectroscopic redshifts of $365$ cluster member galaxies and the total mass is determined as $M_{200,c} = 1.38 \pm 0.25\times 10^{15} M_{\odot}$. The mass estimate based on the X-ray temperature is significantly lower at $M_{200} = 6.20 \pm 0.75 \times 10^{14} M_{\odot}$, providing further indications that merger activity boosts the velocity dispersion and/or biases the temperature low.
An extended X-ray source located south of the field of view also coincides with a galaxy overdensity with spectroscopic redshifts in the range $0.05<z<0.07$. This source further supports the idea that the cluster is undergoing merger activity. Another extended source located north of the field of view is detected in X-rays and coincides with an overdensity of galaxies with spectroscopic redshifts in the range of $0.070<z<0.077$. This is likely a background cluster that is not directly related to A3158. Additionally, the known South Pole Telescope cluster SPT-CL J0342-5354 at $z=0.53$ was detected. } 
{}

\maketitle

\section{Introduction}
As the largest known gravitationally bound objects in the observable universe, galaxy clusters are important for the study of the large-scale structure and cosmology. Galaxy clusters grow through accretion and merging with smaller objects. Clusters typically have masses between $10^{14}$ and $10^{15} M_{\odot}$ and consist of approximately 100 - 1000 galaxies, hot intra-cluster gas, dark matter, and a small population of relativistic particles. Clusters can reach sizes of up to 6 Mpc in diameter and are important astrophysical laboratories for the study of metal abundance, temperature, and gas densities of this hot intra-cluster medium. 
The metal abundance in the outskirts of galaxy clusters, that is, outside $r_{500}$ and inside $3 \sim r_{200}$ \citep{Reiprich_2013}, is particularly interesting. The metal abundance in cluster outskirts is not only useful for estimating the total mass of metals in the current universe, but also retains the information of early enrichment, which reflects the cosmic star formation in early epochs. 
Low-redshift clusters such as Abell 3158 ($z = 0.059$; \citealt{Redshift}) are good places in which to examine the faint outskirts of clusters further. The physical and enrichment processes that are absent or less important in the central regions of clusters, such as minor mergers and infall of gas clumps, likely affect the gas in the cluster outskirts \citep[e.g.][]{Reiprich_2013}. 

The Spektrum Roentgen Gamma (SRG) mission, launched on 13 July 2019 from Baikonur, Kazakhstan, carries two high-energy instruments. The soft X-ray instrument on board this mission is the \textbf{e}xtended \textbf{RO}entgen \textbf{S}urvey and \textbf{I}maging \textbf{T}elescope \textbf{A}rray (eROSITA), which is the most modern X-ray telescope and consists of seven Wolter-1 telescope modules (TMs). The energy range of eROSITA extends from $0.2 - 8.0$\,keV. It will create the first all-sky survey in the X-ray hard band ($2-8$\,keV) and is $\sim 20$ times more sensitive than ROSAT in the $0.2-2$\,keV energy range. eROSITA has a field of view with a diameter of $1.03$ degrees \citep{Predehl_2021, Merloni}. 

During the calibration and performance verification phase (CAL-PV), the Abell 3158 cluster was observed as a calibration target on 21 November 2019 for a duration of 80 ks.  The cluster had previously been observed with the \textit{XMM-Newton} observatory. The appearance of the cluster is relaxed in the profiles from previous studies, making it an excellent candidate on which to perform a cross-calibration of the two telescopes. The Abell 3158 galaxy cluster is described as being undisturbed in \cite{Irwin} and \cite{Lokas}, but
\citet{Hudson} classified the cluster as a non-cool-core cluster due to the lack of a bright central core and a central temperature drop, while
\cite{TempMap} detected an off-centre cool gas clump west of the cluster centre from a 2D temperature map. They speculated that the cool gas clump may be the result of a major merger event. The cool gas clump was accompanied with a bow-shaped edge, and they determined that this cool gas clump moves adiabatically with this edge. The mass of the cluster has frequently been estimated with different methods. \cite{OmegaWINGS} calculated a mass of $1.79 \times 10^{15} M_{\odot}$ using the $M_{200}$ - $\sigma_{v}$ relation from \cite{Finn05}. \cite{AngLiu} estimated the mass from the temperature obtained in X-rays with the mass-temperature (M-T) relation, while \cite{MCXC}  estimated the mass using the mass-luminosity (M-L) relation. The hydrostatic mass was determined as well (e.g. \citep{Chen,X-COPMass}). The values for the mass that were obtained with these methods are typically much lower than those obtained with velocity dispersions. Additionally, the field of A3158 contains a number of extended sources. They have been observed with X-COP \citep{Eckert17}, but are not described in the literature. These extended sources may hold interesting information about the state of the cluster. 

In this work, we analyse eROSITA CAL-PV data of Abell 3158 and compare our results with archival \textit{XMM-Newton} and \textit{Chandra} data, in which the 1D temperature, abundance, and normalisation profiles agree well between the telescopes. The data reduction is described in \S \ref{Section:DataReduction} . In \S \ref{Section:Imaging} and \S \ref{Section:Spectral} we describe the imaging and spectral analysis methods and model components. The galaxy velocity dispersion analysis is outlined in \S \ref{Section:VD}. We present and discuss our results in \S \ref{Section:Results} and conclude with a summary in \S \ref{Section:Conclusion}. 

The cosmology we assumed is $\Lambda$CDM with $\Omega_{\Lambda} = 0.7$, $\Omega_{m} = 0.3,$ and $H_{0} = 70$ km s$^{-1}$ Mpc$^{-1}$. $R_{500} = 1.07$ Mpc is taken from \cite{MCXC}, and using the relation in \cite{Reiprich_2013} of $r_{500} \approx 0.65  r_{200}$, $r_{200}$ is calculated to be $1.64$ Mpc. At the redshift of the A3158 galaxy cluster, these distances translate into $r_{500} = 15.58$\,arcmin and $r_{200} = 23.95$\,arcmin, which are the starting values we used to determine the extraction radii. 

\section{Data reduction}\label{Section:DataReduction}
\label{reduc}
The Abell 3158 observation ID is 700177, and the processing version we used for this work was c001. All seven TMs were operating nominally for the observation. The telescope was pointed for 80\,ks to the direction of the cluster. However, during the observation, each TM carried out an observation for which the  filter wheel was closed (FWC) that lasted between $10 - 15$\,ks to monitor the particle-induced background. Due to this lost observing time, the data with which science can be carried out has an observing time of approximately 60\,ks. 

The eROSITA data shown here were processed using the extended Science Analysis Software System (eSASS) developed by the German eROSITA consortium. The eSASS version that is used for the data reduction processes is the eSASSusers\_201009 user release from October 2020 \citep{2022A&A...661A...1B}.

\subsection{Filtering flares}
\begin{figure}[h]
    \centering
    \includegraphics[trim=10 0 30 20,clip,width=\columnwidth]{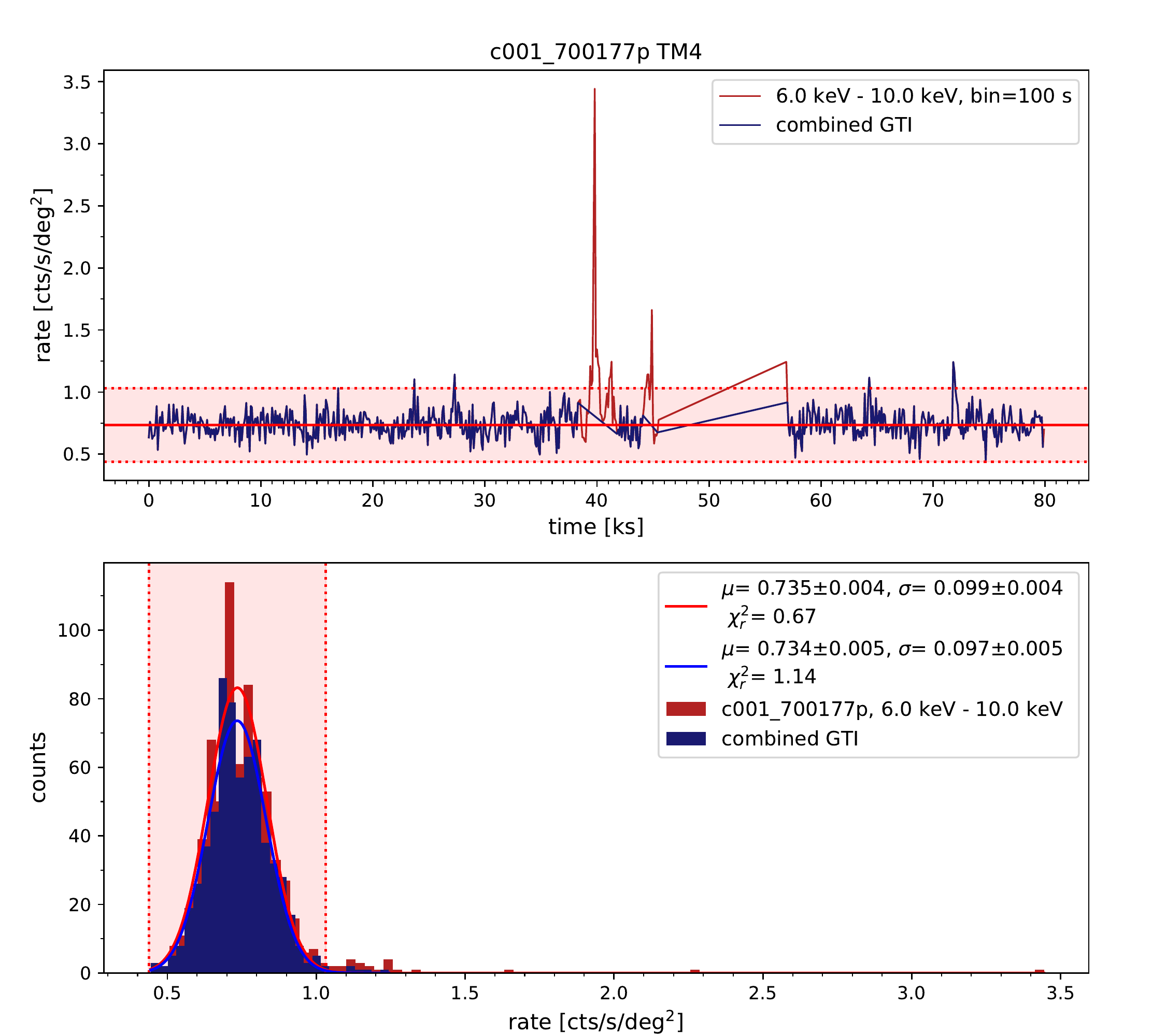}
    \caption{Light curve of TM4 showing a flare that was excluded from the data.}
    \label{fig:Lightcurve4}
\end{figure}

Upon retrieving the data, a flare filtering script was run on the data, a detailed description of which can be found in \cite{A339195}. This was carried out in order to identify any soft proton flares that may have occurred during the observing time. A small number of time bins that have more than $3\sigma$ counts remain as these are likely statistical peaks and not real flares. The light curve of TM4 shown in Fig. \ref{fig:Lightcurve4} shows a flare during the time period $37-42$\,ks of the observation followed by the FWC observation. This flare is present after pattern selection and flagging were applied. The time span of the flare and the FWC period are excluded from the GTI of the observation. The light curves of the remaining telescope modules are included in the appendix.

\begin{figure}
    \centering
    \includegraphics[width=\columnwidth]{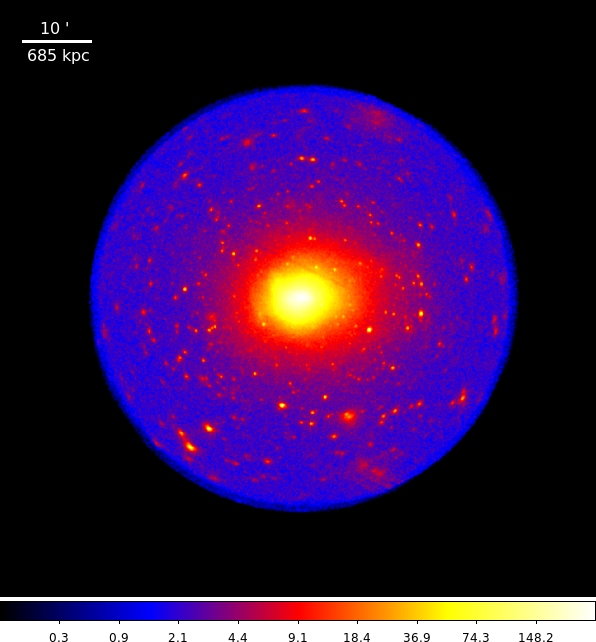}
    \caption{Smoothed photon image created in the $0.2-10$\,keV energy range using the cleaned TM0 event list. TM0 is the combination of all seven telescope modules. The colour bar shows the number of counts.}
    \label{fig:c001img}
\end{figure}

\begin{figure}
    \centering
    \includegraphics[width=\columnwidth]{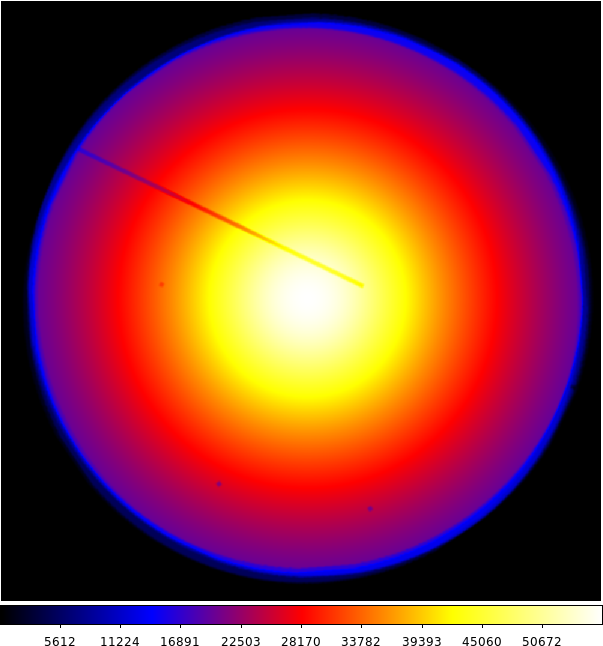}
    \caption{Exposure map created with the energy range $0.3-2.3$\,keV for TM8 and $1.0-2.3$\,keV for TM9.}
    \label{fig:CORRExp}
\end{figure}

Following the cleaning of the data from flares, the GTI of the event lists were updated using \texttt{evtool}. The \texttt{evtool} task can also be used to merge event lists. This feature was used, and a clean merged event list was created of TM0, which is the combination of all seven telescopes. After this, images, exposure maps, and detection masks were created using the eSASS tasks \texttt{evtool}, \texttt{expmap,} and \texttt{ermask} for each telescope and the merged event list. The raw TM0  merged image is shown in Fig. \ref{fig:c001img}. There is an artificial stripe slightly north of the cluster centre. This is caused by a column of bad pixels that was removed from TM2. The default exposure map creation does not take into account bad pixels that are identified for the observation. The effect from the removed column in TM2 was propagated to the exposure map, which is shown in Fig. \ref{fig:CORRExp}. This exposure map was created with the energy range $0.3-2.3$\,keV for  TM1, TM2, TM3, TM4, and TM6 (collectively, TM8) and with the energy range $1.0-2.3$\,keV for TM5 and TM7, which  do not have an on-chip optical blocking filter (collectively, TM9). 

\subsection{Source detection}

Source detection was performed as described in \citet{2021arXiv210614544I}.
All point sources were masked, as was an extended source south of the main cluster. This source was identified as a cluster from the South Pole Telescope catalogue, SPT-CL J0342-5354 \citep{SPT}. This is shown in Fig. \ref{fig:PIBSub}.

\subsection{\textit{XMM-Newton}}
Two pointed observations centred on the A3158 galaxy cluster are available on the \textit{\textbf{X}MM-Newton} \textbf{S}cience \textbf{A}rchive (XSA). The shorter observation is highly contaminated by soft proton flares and was therefore excluded from this analysis. Four pointings of the outskirts of the cluster are located north, west, south, and east of the central pointings. The observation IDs and observing times of each observation are listed in table \ref{tab:XMMObs}.
\begin{table}[h!]
    \centering
    \begin{tabular}[c]{|c|c|c|}
    \hline
    Observation name & Observation ID & Observing time (s)\\
    \hline \hline
    Long & 0300210201 & 22392 \\
    \hline
    Short & 0300211301 & 9408 \\
    \hline
    South & 0744411501 & 31000 \\
    \hline
    West & 0744411401 & 33400 \\
    \hline
    North & 0744411301 & 31399 \\
    \hline
    East & 0744411601 & 33800 \\
    \hline
    \end{tabular}
    \caption{Observation IDs and observing times of the Abell 3158 field with \textit{XMM-Newton}.}
    \label{tab:XMMObs}
\end{table} 
In order to compare the profiles from eROSITA with \textit{XMM-Newton}, the event files were obtained from the archive to be analysed. The analysis carried out in this work closely follows that described in \cite{Ramos-ceja}. To summarise, flare filtering  was carried out in order to identify and remove any soft proton flares that were present in the observation, CCDs that were in an anomalous state during the observation were removed, and the instrumental background and exposure were corrected for. Point sources were identified so that they could be masked during the spectral extraction process. The central coordinates of the cluster were then determined to be R.A. $55.7108$ and Dec. $-53.6304$ from the emission-weighted centre of the \textit{XMM-Newton} central pointing. These central coordinates are used in the analysis of XMM-Newton and eROSITA throughout this paper.
\subsection{\textit{Chandra}}
We used \textit{Chandra} observation IDs 3201 and 3712 for the analysis. The two observations were taken on ACIS-I. The data reduction was performed using the software {\tt CIAO v4.12}, with the latest version of the \textit{Chandra} calibration database ({\tt CALDB v4.9}).
Time intervals with a high background level were filtered out by performing a 3$\sigma$ clipping on the light curve in the $2.3-7.3$\,keV energy range and were binned with a time interval of $200$\,s. The cleaned exposure times were $24.5$\,ks and $27.3$\,ks for 3201 and 3712, respectively. Point sources within the ICM were identified with {\tt wavdetect} and were masked after visual inspection. The ancillary response file (ARF) and redistribution matrix file (RMF) were computed using the commands {\tt mkarf} and {\tt mkacisrmf}. Because the emission of A3158 covers the whole CCD area, we extracted and processed the background from the ``blank sky" files using the {\tt blanksky} script.

\section{Imaging analysis}\label{Section:Imaging}
\subsection{Subtraction of particle-induced background}
\begin{figure}
    \centering
    \includegraphics[width=\columnwidth]{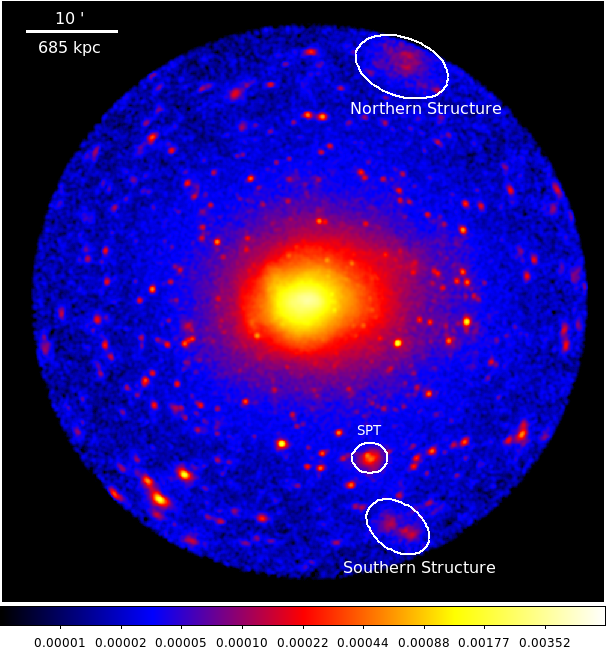}
    \caption{Count-rate image with PIB-subtraction in the $0.3-2.3$\,keV energy band. The extended source south of the cluster was identified as the SPT cluster.}
    \label{fig:PIBSub}
\end{figure}

A detailed description of the subtraction  of particle-induced background (PIB) is provided in \cite{A339195}. Using the FWC data reprojected to the Abell 3158 direction, the same method was used in this work to create a background-subtracted image in the $0.3-2.3$\,keV energy band. The process was carried out in the $1-2.3$\,keV energy band for
TM5 and TM7. The PIB-subtracted photon image was divided by the combined exposure map to create a PIB-subtracted count-rate image. The product of this process is shown in Fig. \ref{fig:PIBSub}.
\subsection{Surface brightness profile}
A region file was created with annuli in steps of 10 arcseconds spanning the area from the centre of the cluster to the $r_{200}$, masking the point sources. With ftools, this region file was used to determine the number of counts in the PIB-subtracted photon image that was created in the previous step. The exposure map that was created in this process was also used in order to determine the exposure time in these annuli. An annulus from $r_{200}$ out to $31.00$\,arcmin was determined as the sky-background region, and the number of counts and the exposure time in this region were extracted using the same method as described above. 

\begin{figure}
    \centering
    \includegraphics[width=\columnwidth]{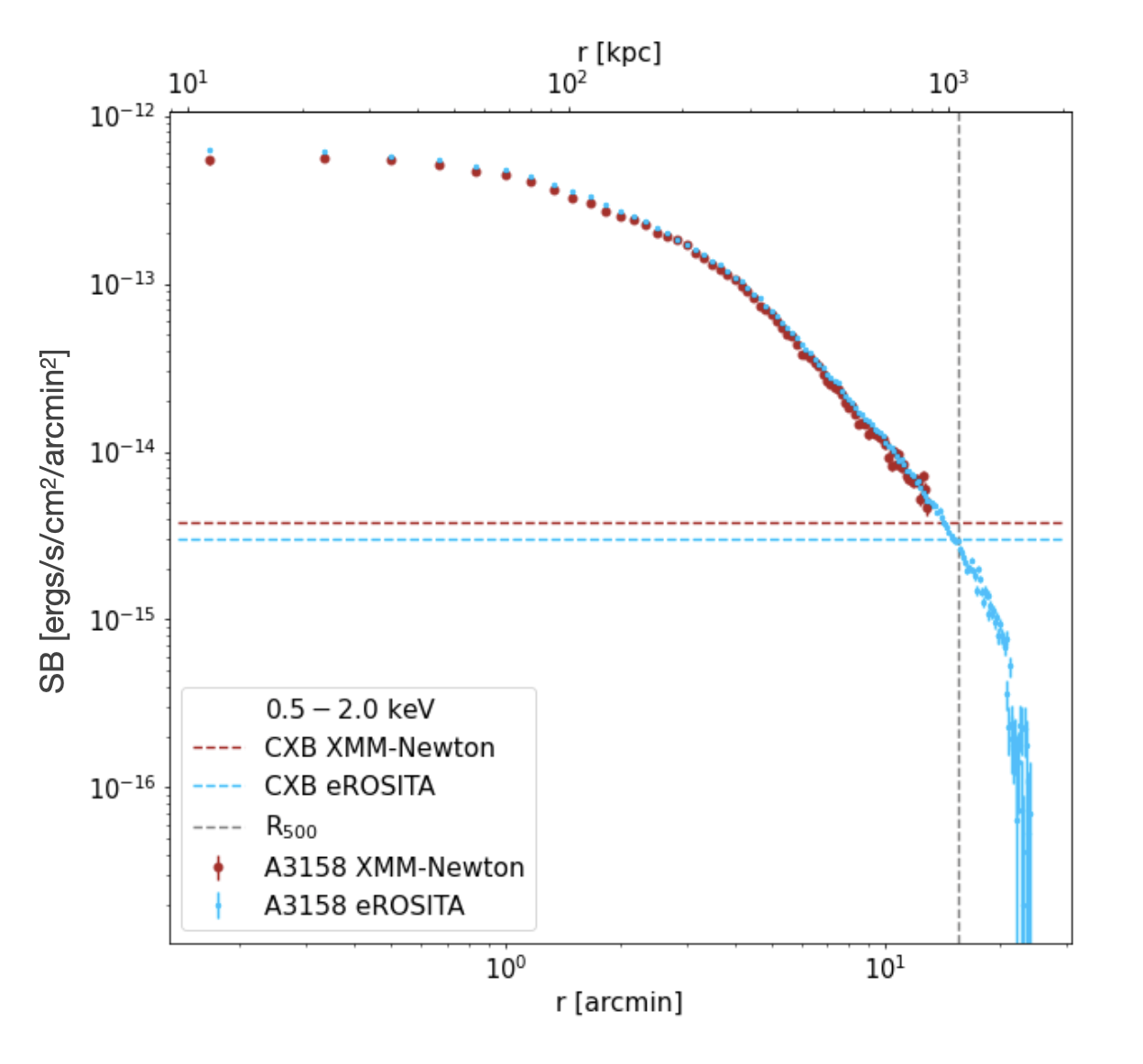}
    \caption{Comparison of the surface brightness profile from XMM-Newton out to $r_{500}$ and from eROSITA out to $r_{200}$. The data points are PIB subtracted and sky-background subtracted.}
    \label{fig:surfB}
\end{figure}

The surface brightness for each annulus was calculated by dividing the counts by the exposure and the area and subtracting surface brightness of the background region. This surface brightness was then plotted as a function of distance from the centre of the cluster. The surface brightness profile is shown in Fig. \ref{fig:surfB}. The error bars shown in the profile are determined from the PIB-subtracted photon image, the exposure map, and the sky background. These errors were propagated through the calculations and were taken into account when the final value of the surface brightness of each region was determined. 

\begin{figure}
    \centering
    \includegraphics[width=\columnwidth]{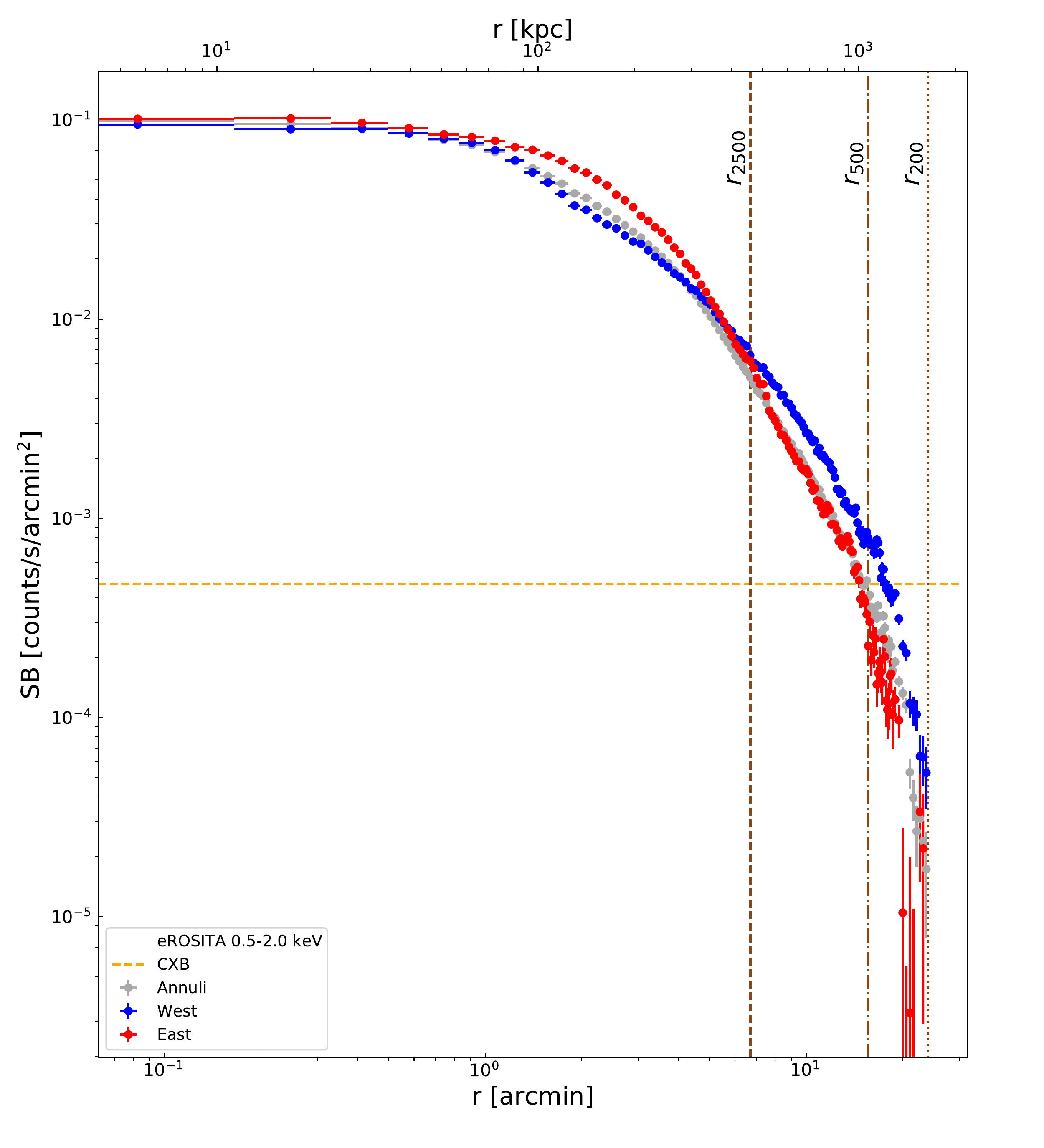}
    \caption{eROSITA clean SB profiles of the full azimuthal (grey), western (blue), and eastern (red) directions out to the $R_{200}$ of the cluster. The CXB level is plotted as the dashed orange line. The characteristic radii are indicated by vertical brown lines.}
    \label{fig:eROsurfB}
\end{figure}
A surface brightness profile was created using the \textit{XMM-Newton} data with the same method as described above. The sky background was determined from the background spectral fitting region in the four outer pointings.

Additionally, to investigate the morphology of the cluster, we calculated and compared eROSITA surface brightness profiles in the western and eastern directions 90 deg opening angle each. The profiles are presented in Fig.~\ref{fig:eROsurfB}.

\section{Spectral analysis}\label{Section:Spectral}
\subsection{Profiles}\label{Section:Profiles}
\label{T-profs}
Using the \texttt{srctool} task in eSASS, spectra were extracted for 13 annuli covering up to $r_\mathrm{200}$, centred at the position (R.A.,Dec.)=($55.7108^{\circ}$,$-53.6304^{\circ}$). In the process, \texttt{srctool} also extracted the necessary files to perform spectral fitting, that is, the auxiliary response file (ARF) and the redistribution matrix file (RMF). In addition, we extracted spectra for a large outer bin from 23 to 31 arcmin, which we used to constrain the local properties of the X-ray background. All spectra were then grouped with a minimum of $25$ counts in each channel.

For all of the spectral fitting carried out in this paper, the {\tt XSPEC 12.11.1}
software package \citep{XSPEC} was used. We estimated the thermal emission from optically thin plasma with version 3.0.9 of the \texttt{APEC} model, using the solar abundance tables of \citet{2009Asplund} as reference. The \texttt{tbabs} model \citep{2000Wilms} was used throughout to compute the attenuation due to photoelectric absorption, assuming a fixed Galactic hydrogen column density, $N_{H}$, of $1.4\times 10^{20}$ cm$^{-2}$. This value was retrieved from the UK Swift Science Data Centre\footnote{\url{https://www.swift.ac.uk/analysis/nhtot/index.php}} and relies on the method of \cite{Willingale}.

The 14$\times$7 spectra from all annuli and TMs were loaded and fitted simultaneously to a model consisting of the instrumental background, the X-ray background and an \texttt{APEC} thermal model at redshift $z=0.059$ \citep{Redshift} for the cluster emission.
The instrumental background was described by the TM-dependent models provided with the eROSITA Early Data Release\footnote{\url{https://erosita.mpe.mpg.de/edr/eROSITAObservations/EDRFWC/}} - a combination of a double broken power law with two additional power laws to describe the continuum, together with several instrument lines, all constrained from the FWC observations. All the parameters describing the the continuum components were fixed and their normalisations were tied to the detector area, but a global renormalisation factor was allowed to vary for the model of each TM. This is motivated by the constant spectral shape observed across FWC datasets. The normalisations of all instrumental lines, however, were left free since the statistics of the FWC data were not sufficient to constrain their spatial and temporal variability in detail. The X-ray background was modelled with an unabsorbed $0.099$\,keV thermal \texttt{apec} \citep{APEC} component to model the Local Hot Bubble (LHB) emission, an absorbed $0.22$\,keV thermal \textit{apec} component to model the Milky Way halo (MWH) \citep{MWH} emission, and an absorbed power-law component with a photon index of $1.41$ \citep{DeLuca} to model unresolved emission from active galactic nuclei (AGN). The metal abundance and redshift of both thermal components were frozen to $Z = 1.0$ $Z_{\odot}$ and $z = 0.0$. The normalisation of each component was left free to vary, but was tied across annuli and TM in proportion of the covered sky area.

The fit was carried out in the $0.5-9.0$\,keV energy range for telescope modules with an on-chip filter, while the fit for the telescope modules without an on-chip filter was carried out in the $0.8-9.0$\,keV energy range. The data of the large outer annulus were fitted first to constrain the X-ray background, allowing for a residual contribution from the cluster outskirts. Then all other annuli were added and fitted together. The temperature, abundance, and normalisation for the cluster were varied while the model was fit. The statistics used in the fitting process was the Poisson-distributed C-statistic \citep{Cash1979}. The best-fit values of the instrumental background normalisations were consistently higher than the reference FWC models, by values ranging from 1.3\% (TM4) to 5.8\% (TM1). This level is compatible with the variability of about 6\% in normalisation that is observed across FWC datasets. In order to ensure that these results are robust, we also attempted to fit the data with $\chi^2$ statistics and a minimum grouping of $100$ counts in each channel. The results obtained using the two methods were very similar, and the C-statistic was therefore used for the analysis.
We also confirmed the distribution of the C-statistic for all bins with the steppar command, which appeared well behaved and confirmed the robustness of both the best-fit values and the estimated error ranges.

\subsection{XMM-Newton}
The spectral extraction of the source spectra was carried out just with the central pointed observation of A3158, and the sky-background spectra were extracted in the same area as eROSITA from four surrounding pointed observations that were part of the \textit{\textbf{X}MM-Newton} \textbf{C}luster \textbf{O}utskirts \textbf{P}roject (X-COP) \cite{Eckert17}. The spectra were fit with the same model as described in \cite{Ramos-ceja}. This model is similar to the model that was fit to the eROSITA spectra with the inclusion of a number of instrumental lines, and $\chi^2$ statistics were used in the fitting process. The fit was carried out in the same energy range as for the eROSITA analysis.

\subsection{\textit{Chandra}}
Limited by the field of view (FoV), we only obtained the profiles within 9\,arcmin.
The spectra were fitted
adopting the C-statistic.
The full band (0.5--7~keV) spectrum was fitted with a single {\tt apec} 
model.
The redshift was fixed, but the temperature, metal abundance, and normalisation were set as free parameters.

\subsection{Temperature map}\label{Section:TmapMethod}
\cite{TempMap} performed a thorough X-ray analysis of the Abell 3158 galaxy cluster with \textit{XMM-Newton} and \textit{Chandra} observations. A 2D temperature map of the cluster was created, and an off-centre cool gas clump was highlighted in the paper. 
According to these authors, based on the surface brightness distribution, A3158 had previously been described as a regular cluster; nonetheless,
it does not host a cool core that relaxed clusters are typically expected to have. Due to this previously discovered substructure and the apparent lack of a cool core, we decided to create a temperature map with the new eROSITA data. The contour-binning package from \cite{contbin} was used for this process. 

The outer region used to characterise the X-ray background (r$>$23\,arcmin, Section~\ref{T-profs}) was
excluded from the creation of the bin map. 
Two masks were created to be used as input into the contour-binning script from \cite{contbin}. The first mask we created excluded the observation outside $r_{500}$ , and the second mask excluded the observation inside $r_{500}$ and outside $r_{200}$. A signal-to-noise ratio of $150$, a constraining value of $1.5$, and a smoothing signal-to-noise ratio of $50$ were implemented on the first mask. A signal-to-noise ratio of $200$ was implemented on the second mask. The two resulting bin maps were merged to create one final bin map with larger bins in the outer radius. 

An extraction code that used the \texttt{srctool} task from eSASS was implemented. This code took the bin map and a weight image as inputs and created for each bin a mask and a weighted mask. This mask was used as the input for  \texttt{srctool,} and the spectra, ARF, and RMF were extracted. After the spectra were extracted, the FWC data were renormalised, and the spectra were grouped with at least $\text{five}$\,counts in each bin.

The fitting script relied on the methods described in Section~\ref{T-profs}, with the notable difference that each cell of the temperature map was fitted separately, in combination with the outer annulus used to constrain the background. 
First, the fit was carried out with the abundance fixed at $0.3$, after this, the abundance was varied and the fit was run again.

The products of the script are a temperature map with a fixed abundance, a temperature map with a free abundance, and an abundance map. Bins with temperatures higher than $30.0$\,keV, a reduced $\chi^2$ value for the fit greater than $1.5,$ or a relative error value greater than $50\%$ were not included in the temperature map with the fixed abundance. The same criteria were used for the temperature map with a varying abundance. An added criterion was that abundance values with a maximum error value that was lower than the measured value were rejected. The temperature map with the free abundance is shown in \S \ref{Section:Results}.

\subsection{Mass-temperature relation}
Using an annulus extending from $0.2-0.5$ $r_{500}$, we extracted the spectra and determined the temperature in this region following the fitting model described above. The temperature measured in this region was $5.158\ \pm\ 0.038$\,keV. From this measurement of the temperature $T_{500}$, the M-T scaling relation $$ \log\left(\frac{M_{500}}{C_1}\right) = a \cdot \log\left(\frac{T_{500}}{C_2}\right) + b  $$ from \cite{Scaling} was used to estimate the cluster mass, where $C_{1} = 5\times10^{13}h^{-1}_{70} M_{\odot}$ and $C_{2} = 2$\,keV are constants of the scaling relation, and $a = 1.62 \pm 0.08$ and $b = 0.24 \pm 0.04$ are the fit results of the scaling relation for HIFLUGCS cluster with a temperature $kT > 3$\,keV. The mass value was determined as $M_{500} = 4.03 \pm 0.49 \times 10^{14} M_{\odot}$. The uncertainties of the scaling relation was propagated here. The value for the $r_{500}$ was then estimated to be $1.10 \pm 0.04$ Mpc according to the relation described in \cite{Reiprich_2013}. The relation between $r_{500}$ and $r_{200}$ of $r_{500} \approx 0.65 r_{200}$, which assumes an NFW \citep{Navarro} profile with concentration $c = r_{200}/r_{s} = 4$ was then used to estimate $M_{200}$ , which was calculated to be $M_{200} = 6.20 \pm 0.75 \times 10^{14} M_{\odot}$. This value for the mass is compared to the mass determined with an $M_{200}$ - $\sigma_{v}$ relation in \S \ref{section:Mass}. 

\section{Velocity distribution and redshifts}\label{Section:VD}

\subsection{Member galaxies}
Following the OmegaWINGS galaxy spectroscopy survey \citep{OmegaWINGS}, we identified member galaxies in the Abell 3158 field using a redshift cut of $0.05<z<0.07$, a magnitude cut of $20$ mag in the V filter, and a spatial limitation of the galaxies within $30$\,arcmin of the cluster.
Implementation of these cuts in a search using the NASA/IPAC Extragalactic Database (NED)\footnote{\url{http://ned.ipac.caltech.edu/}}
{results in} a total number of 365 member galaxies.
This list contains positions and spectroscopic redshifts {from the following} surveys: \cite{2MASX78}, \cite{LDM95}, \cite{ESO98}, \cite{APMUKS04}, \cite{WINGS09}, and \cite{OmegaWINGS}. 

\begin{figure}
    \centering
    \includegraphics[width=0.8\columnwidth]{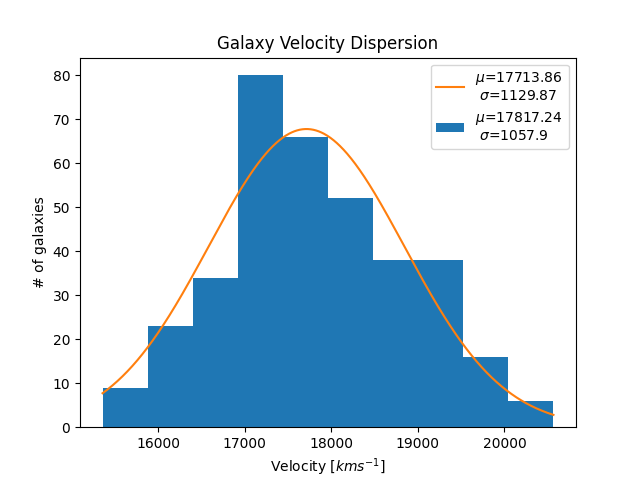}
    \caption{Histogram of member galaxy velocities with a redshift range of $0.05<z<0.07$. The orange curve is the best fit Gaussian of the data set.}
    \label{fig:Hist1}
\end{figure}

Using the redshift measurements, we calculated a rudimentary velocity using $v=cz$, where $c$ is the speed of light in vacuum. A histogram of the galaxy velocities is plotted in Fig. \ref{fig:Hist1}. From this histogram, we determined the standard deviation of the data set of $1058 \pm 41$ kms$^{-1}$ as the velocity dispersion of the cluster member galaxies. 
The errors on the redshifts were obtained from \cite{OmegaWINGS} and from the NED. The errors on the velocities were calculated based on this. The error of the velocity dispersion was determined to be the standard deviation of the error values of the velocities. 

\begin{figure}
    \centering
    \includegraphics[width=\columnwidth]{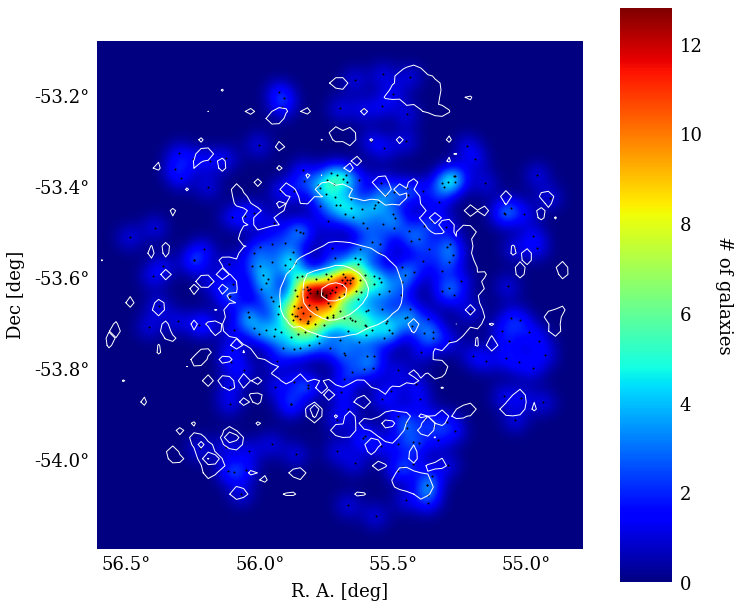}
    \caption{Galaxy density map with members of Abell 3158. The extended source in the south of field coincides with a small overdensity of galaxies, the location of which can be seen in Fig. \ref{fig:PIBSub}. These galaxies have redshifts in the range $0.05<z<0.07$.}
    \label{fig:DM1}
\end{figure}

Along with the histogram of galaxy velocities, using the positions of the member galaxies, a density map of the list of member galaxies was created with X-ray contours from eROSITA overlaid. This map is shown in Fig. \ref{fig:DM1}.

The velocity dispersion measured was then used to determine the cluster mass. \cite{OmegaWINGS} used the  $M_{200}$ - $\sigma_{v}$ relation from \cite{Finn05}. With a $\sigma_{v}$ value of $1023$ kms$^{-1}$ , they calculated $M_{200,c}$ as $1.79 \times 10^{15} M_{\odot}$. 

Using the same method, we determined $M_{200,c}$  with the velocity dispersion determined in this work ($1058 \pm 41$\,kms$^{-1}$). The mass is calculated to be $1.98 \pm 0.23 \times 10^{15} M_{\odot}$ , and the errors were determined by Gaussian error propagation. The value of the mass determined in this paper is consistent with the mass reported in \cite{OmegaWINGS}. 

Implementation of an updated $M_{200} - \sigma_{v}$ relation from \cite{Bocquet} with the velocity dispersion from both this work and \cite{OmegaWINGS} results in  $M_{200,c}$ values of $1.38 \pm 0.25 \times 10^{15} M_{\odot}$ and $1.23 \pm 0.19 \times 10^{15} M_{\odot}$ , respectively. This shows that our results agree with those of \cite{OmegaWINGS}, as expected given the consistent velocity dispersion, although the $M_{200,c} - \sigma_{v}$ relations from \cite{Finn05} and \cite{Bocquet} are not consistent.

\begin{figure}
    \centering
    \includegraphics[width=\columnwidth]{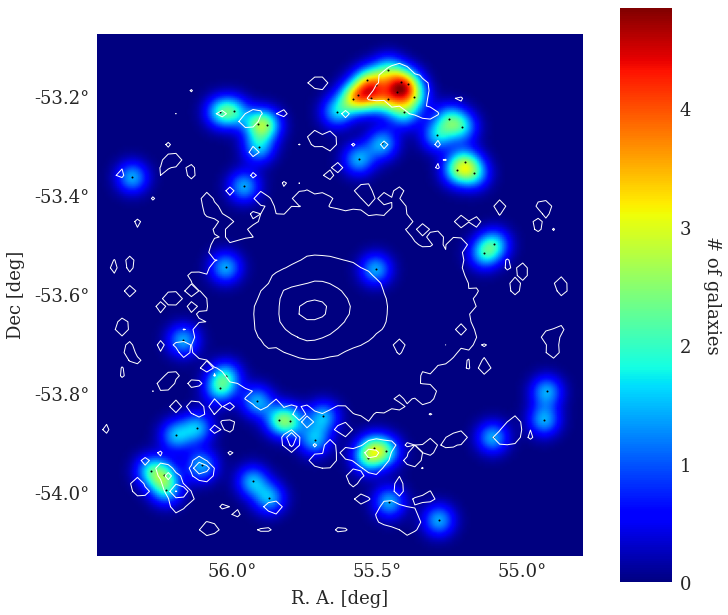}
    \caption{Galaxy density map of member 2 galaxies that are classed as probable separate substructure. These galaxies have a redshift in the range $0.070<z<0.077$. X-ray contours are overlaid and show that the extended source in the north is located in the same region as the galaxy overdensity.}
    \label{fig:DM2}
\end{figure}

The OmegaWINGS survey contains a number of galaxies that have been identified as probable substructure members. These galaxies all have a redshift value in the range $0.070<z<0.077$.
A galaxy density map
with galaxies in this redshift range
is shown in Fig.~\ref{fig:DM2}. 

\begin{figure}
    \centering
    \includegraphics[width=0.85\columnwidth]{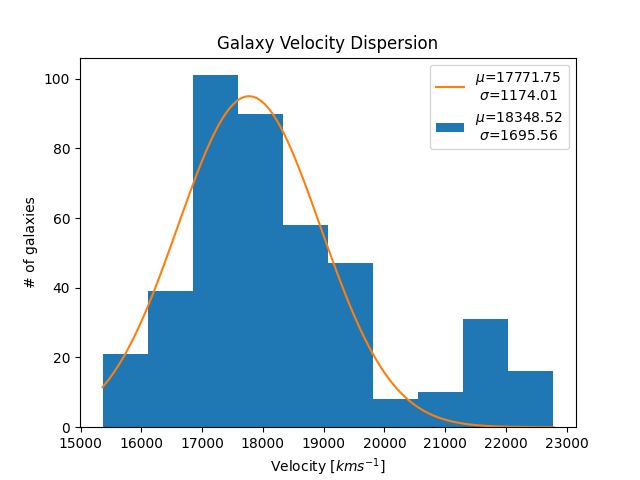}
    \caption{Histogram of galaxy velocities with members of A3158 and members of a separate substructure. The substructure can be seen as an increased population of velocities right of the initial peak. The orange curve is the best-fit Gaussian to the data set.}
    \label{fig:Hist2}
\end{figure}

Combining the list of members of A3158 and the list of substructure members, we plotted a velocity histogram. This is shown in Fig. \ref{fig:Hist2}. 

\subsection{Serendipitous extended sources}
Three very obvious extended sources were observed serendipitously in the Abell 3158 pointed observation. Two of them are located south of the cluster, and one is located north or north-west of the cluster, as indicated by the white ellipses and circle in Fig.~\ref{fig:PIBSub}. The extended source that is closest to the cluster centre south of the field of view was identified as a $\sim 0.5$ redshift cluster that was previously observed by the \textbf{S}outh \textbf{P}ole \textbf{T}elescope (SPT) survey \cite{SPT}, SPT-CL J0342-5354.

While this cluster is detected with very high significance in the eROSITA observation, the number of source photons is insufficient for a detailed spectroscopic analysis that could constrain an X-ray redshift. 
Furthermore, no galaxies with a spectroscopic redshift close to that of the cluster were observed in the optical band. In Fig. \ref{fig:DM2}, it appears that an overdensity of galaxies in the redshift range $0.070 < z < 0.077$ coincides with an X-ray overdensity with the same location as the SPT-CL J0342-5354 cluster. We speculate that this is a coincidence. 

The two remaining extended sources in the field of view have not been documented previously. As the sources are at the edge of the field of view, the count rates are low and a spectral analysis is challenging.
However, the galaxy density maps in \S \ref{Section:VD} show overdensities of galaxies that overlap with the X-ray overdensities. The locations of these extended sources are listed in table \ref{tab:RADecTab}. The unabsorbed X-ray luminosity of these structures was also estimated through the conversion of background-subtracted count rates
in the energy band of $0.5-2.0~\mathrm{keV}$. We used an \texttt{apec} component, assuming gas with a temperature of 1 keV and an abundance of $0.3Z_\odot$. For the northern structure, the $L_X$ was calculated from a 3.75 arcmin radius region centred at the X-ray emission peak listed in Table \ref{tab:RADecTab}. For the southern structure, we defined a 3.25 arcmin radius centred at (R.A., Dec.) = (55.4300, -54.0502) in order to best capture its entire larger-scale emission, given its substructured nature. 

 The resulting luminosities correspond to those of very low-mass galaxy groups. They correspond for instance to the lowest-luminosity systems in the eeHIFLUGCS-like sample studied in \citet{2020A&A...636A..15M}.

\begin{table}[h!]
    \centering
    \resizebox{\columnwidth}{!}{\begin{tabular}[c]{|c|c|c|c|c|}
    \hline
    \multirow{2}{*}{Location} & R.A. & Dec. & $z$ & $L_X$ \\
     & (J2000) & (J2000) &  range & $[10^{42}~\mathrm{erg\cdot s}^{-1}]$\\
    \hline \hline
    North & 55.4216 & -53.1991 & 0.070 - 0.077 & $3.35-4.82$ \\
    \hline
    South & 55.3949 & -54.0638 & 0.05 - 0.07 & $1.11-2.71$ \\
    \hline
    \end{tabular}}
    \caption{Right Ascension (R.A.) and Declination (Dec.) of the extended sources detected north and south of the FoV. The X-ray luminosity ($L_X$) is calculated in the energy band of $0.5-2.0~\mathrm{keV}$.}
    \label{tab:RADecTab}
\end{table}

\section{Results and discussion}\label{Section:Results}
\subsection{1D profiles}

Fig.~\ref{fig:surfB} shows that the surface brightness profiles determined with XMM-Newton and eROSITA agree well with one another. The eROSITA profile extends to well beyond 20 arcmin. A continuous steepening is observed. Furthermore, at about 20 arcmin, a fairly sharp drop appears. This drop also appears in Fig.~\ref{fig:PIBSub}. On the other hand, the background might also be very slightly oversubtracted because the background annulus starts at $r_{200}$, and this may cause this sharp drop close to $r_{200}$.

Already around 1 arcmin, the surface brightness profiles east and west start to deviate from one another (Fig.~\ref{fig:eROsurfB}). The eastern surface brightness profile (red) drops faster than the western profile (blue) after about 2 arcmin, while at smaller radii, the opposite trend is observed. These two features are consistent with the bow-shaped edge and the gas extension in the west observed in the image (Fig.~\ref{fig:PIBSub}) and further discussed below. A sharp drop at about 20 arcmin is observed for both profiles.

\begin{figure}
    \centering
    \includegraphics[width=1.0\columnwidth,trim=0.5cm 1.0cm 1.2cm 1.0cm,clip]{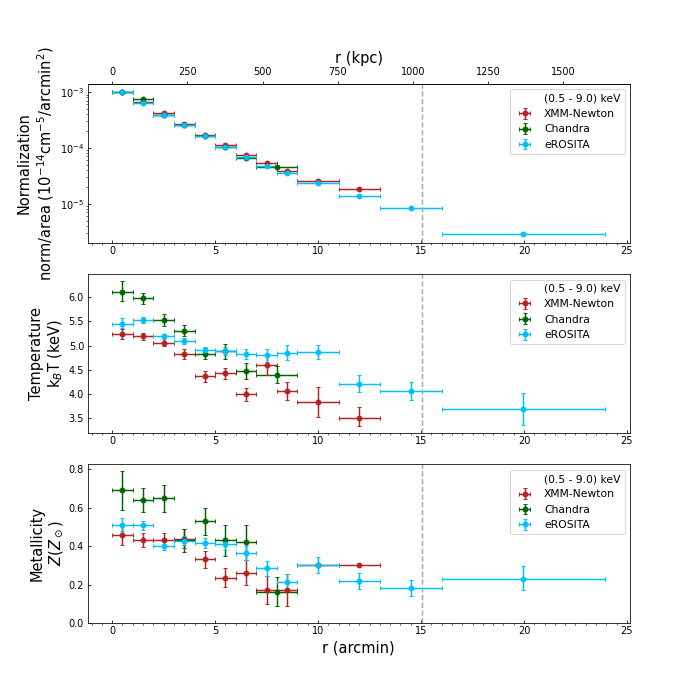}
    \caption{Normalisation, temperature and metal abundance profiles showing the comparison of eROSITA to \textit{XMM-Newton} and \textit{Chandra}. The grey dashed line represents the $r_{500}$ of the cluster.} 
    \label{fig:1DProf}
\end{figure}

\begin{figure}
    \centering
    \includegraphics[width=\columnwidth]{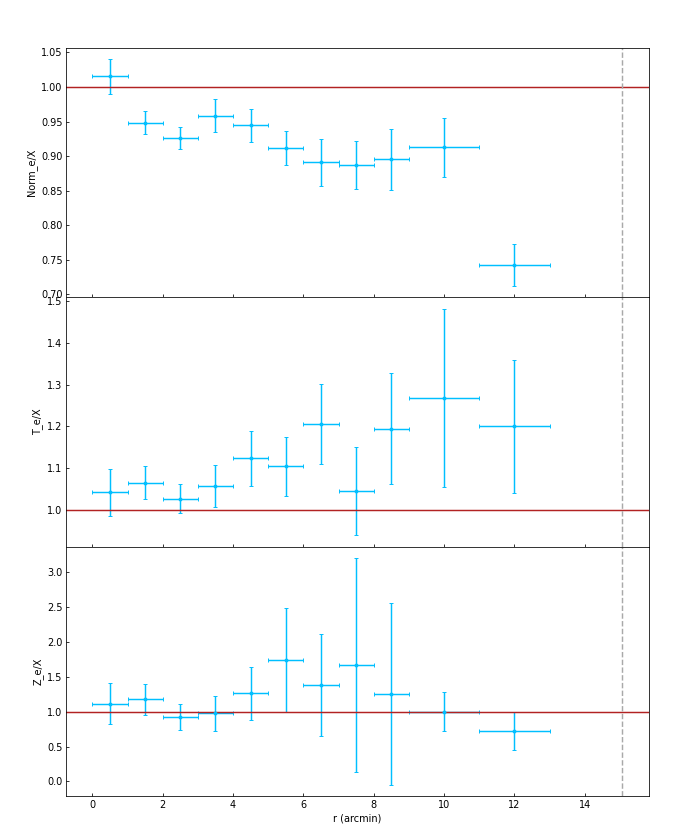}
    \caption{Plot of the normalisation, temperature, and abundance ratios of eROSITA/\textit{XMM-Newton}. The dashed grey vertical line is the $r_{500}$ of the cluster.}
    \label{fig:ratioXMM}
\end{figure}

Following the steps outlined in \S \ref{Section:Profiles}, the temperature, abundance, and normalisation profiles of A3158 were extracted and are shown in Fig. \ref{fig:1DProf}. The temperature profile clearly shows that the measurements from all three telescopes indicate that the cluster does not have a cool core. The results from eROSITA are $\sim 0.5$\,keV lower than the results from \textit{Chandra} in the central region and are slightly higher overall than \textit{XMM-Newton} results out to $r_{500}$.
Although the telescopes disagree with each other, the shape of the temperature profile is similar. The shape also agrees with the published \textit{Chandra}  central temperature profile (\citealt{Hudson}, their Fig.~2).
In the abundance profile, the outer two annuli for the \textit{XMM-Newton} observation are frozen at $Z=0.3$ $Z_{\odot}$ because  these values were estimated to be very low during the fitting and the temperature and normalisation results were affected. The low photon count in the outskirts of the \textit{XMM-Newton} central pointing field of view likely has an impact on this issue. This issue is resolved with eROSITA, and abundance measurements past $r_{500}$ are possible with good constraints.
The normalisation values of \textit{XMM-Newton} are higher by 5\%--10\% than those based on the eROSITA measurements over the full field of view, with the exception of the central and outermost bin. 
This \textit{appears clearly} in the ratio plot shown in Fig. \ref{fig:ratioXMM}. The point spread functions of both instruments are similar.

A comparison of the results from X-COP \citep{Eckert17} and the \textit{XMM-Newton} analysis performed in this work shows agreement in the central regions at a level of $\sim 10\%$  and agreement in the outer regions at a level of $\sim 20\%$ . The larger difference in the outer regions is expected because our analysis of the XMM-Newton data used the X-COP pointings only for the background determination, while \cite{Eckert17} used these pointings to determine both the background and the profiles out to larger radii. The outer X-COP temperature profile agrees very well with our eROSITA results.
eROSITA, however, allows us to constrain the ICM properties out to $r_{200}$ when X-COP stopped at $r_\mathrm{500}$. In addition, the excellent signal-to-noise ratio of eROSITA allows for more radial bins, with a smaller total uncertainty already from radii as small as  $0.5\,r_{500}$, demonstrating the power of eROSITA for cluster outskirts studies if more observations of local clusters at this depth were performed after the completion of the all-sky survey. The relatively high metallicity ($\sim$0.2 solar) well beyond $r_{500}$ may indicate an early (i.e. before cluster collapse) enrichment process dominated by galaxy winds and AGN feedback instead of ram pressure stripping. This has first been described in a cluster system by \citet{2008PASJ...60S.343F} and was later confirmed in other systems as well (see e.g. the recent review by \citealt{2022arXiv220207097M}). With eROSITA, we have now shown that A3158 follows the same trend. This corroborates the high-redshift enrichment scenario.

The ratio plot in Fig. \ref{fig:ratioXMM} also clearly shows that the normalisation, temperature, and abundance profiles generally agree at a  level of $\lesssim 10\%$ . This is also true for the comparison between eROSITA and \textit{Chandra}, the ratio plot of which can be found in the appendix. This is an acceptable level of agreement when compared to the other satellites so shortly after the beginning of eROSITA operations.
Moreover, the presence of multi-temperature structure in the ICM as observed in the temperature map may contribute to differences between X-ray telescopes.
Fitting the spectra with a single-temperature model can lead to a dependence on the observing telescope, although naively, we would expect the opposite trend in temperature \citep[e.g. Fig.~18 in][]{Reiprich_2013}.
 
Additionally, the discrepancy between \textit{XMM-Newton} and \textit{Chandra} has previously been documented for example in \citet[quantitatively confirmed again in \citealt{2020A&A...636A..15M}, their Fig.~A.8]{Schellenberger15} to arise from the systematic effective area calibration uncertainties, and the difference observed between the results from these two telescopes is therefore expected.
Because the eROSITA telescope has a larger field of view than either \textit{XMM-Newton} or \textit{Chandra}, we can extract the profiles out to larger radii overall and deliver good measurements and constraints in the abundance profile, for instance.

\subsection{Cluster morphology}
\label{morpho}
\begin{figure}
    \centering
    \includegraphics[width=0.45\textwidth]{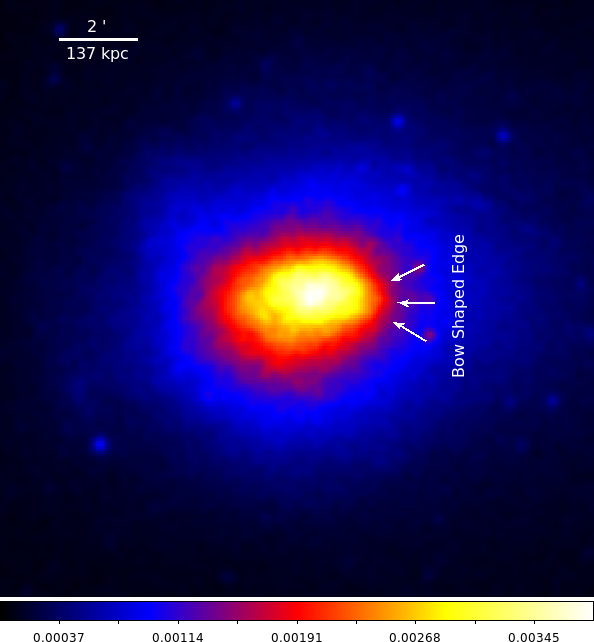}
    \includegraphics[width=0.45\textwidth]{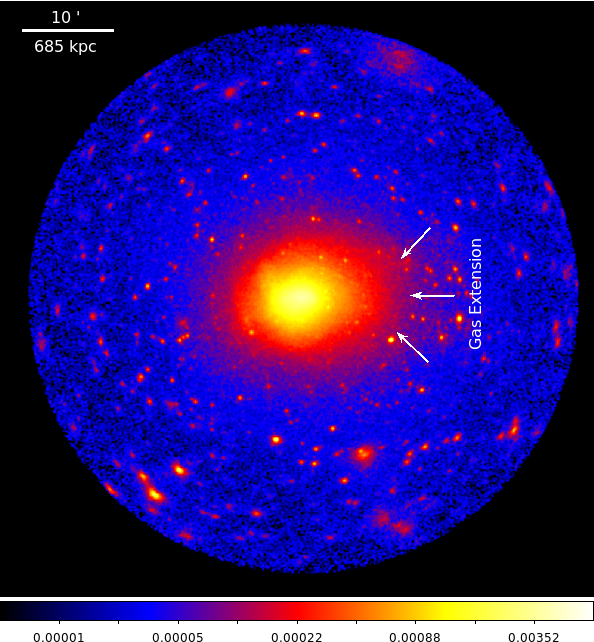}
    \caption{PIB-subtracted count-rate images. Top: image on a linear scale, zoomed in to the central region to show the bow-shaped edge. Bottom: image on a logarithmic scale. The arrows show the large extension of gas west of the cluster. Both images are smoothed.}
    \label{fig:PIBZoom}
\end{figure}

The PIB-subtracted background image is shown
in Fig. \ref{fig:PIBZoom} on two different scales. The top panel is on a linear scale. It is zoomed in to the central region of the cluster. The bow-shaped edge west of the emission peak is close to the location of the cool gas clump discussed in Sect. \ref{Section:TmapDiscussion}. This was observed by \cite{TempMap}, who have determined that the cool gas clump moves adiabatically behind this bow-shaped edge, which was determined to be a faint cold front. This edge exists approximately $ 2$\,arcmin ($\sim 137$ kpc) from the cluster centre. 

Additionally, an extension of gas lies about $ 10$\,arcmin ($\sim 865$ kpc) west of the cluster centre. It is shown in the bottom panel on a logarithmic scale. We present this extension of gas as a new finding. The irregularities between the different scales would suggest that a sloshing effect \citep[e.g.][]{2001ApJ...562L.153M} might occur in the cluster. This further supports the claim that the cluster is undergoing merger activity. 

\subsection{Temperature map}\label{Section:TmapDiscussion}
\begin{figure}
    \centering
    \includegraphics[width=\columnwidth]{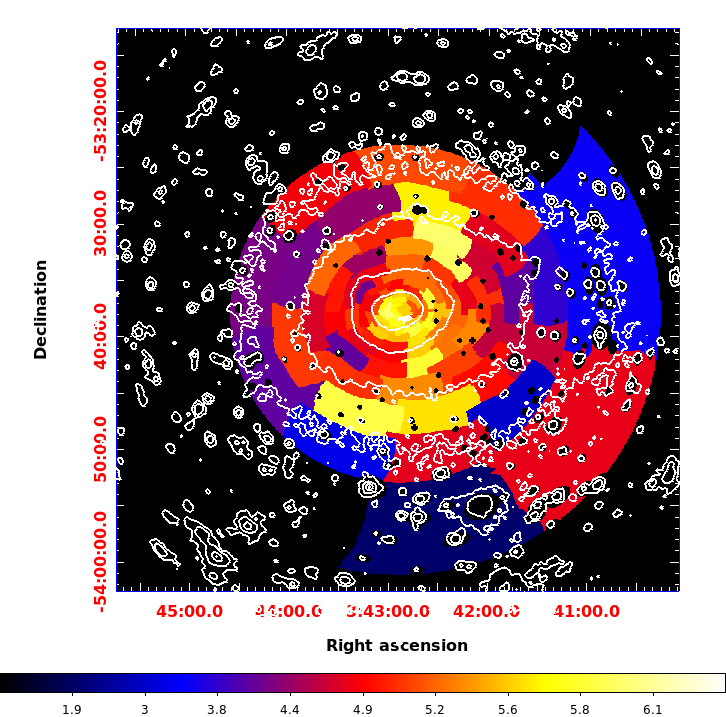}
    \caption{Temperature map of Abell 3158 from eROSITA. The SPT cluster has been masked. The colour bar represents the temperature in keV. eROSITA A3158 contours are overlaid in white.}
    \label{fig:Tmap}
\end{figure}

The temperature map that was created using the steps in Sect. \ref{Section:TmapMethod} is shown in Fig. \ref{fig:Tmap}. This temperature map immediately clearly indicates the cluster does not have a cool core, as would be expected in a regular-looking cluster such as this. 
\begin{figure}
    \centering
    \includegraphics[width=\columnwidth]{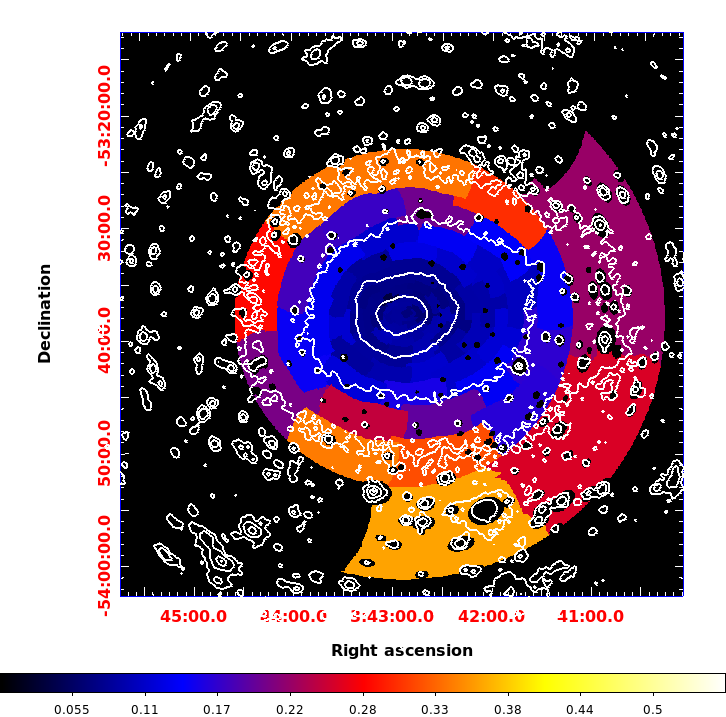}
    \caption{Relative error map corresponding to figure \ref{fig:Tmap}. eROSITA A3158 contours are overlaid in white.}
    \label{fig:Errmap}
\end{figure}
\begin{figure}
    \centering
    \includegraphics[width=0.95\columnwidth]{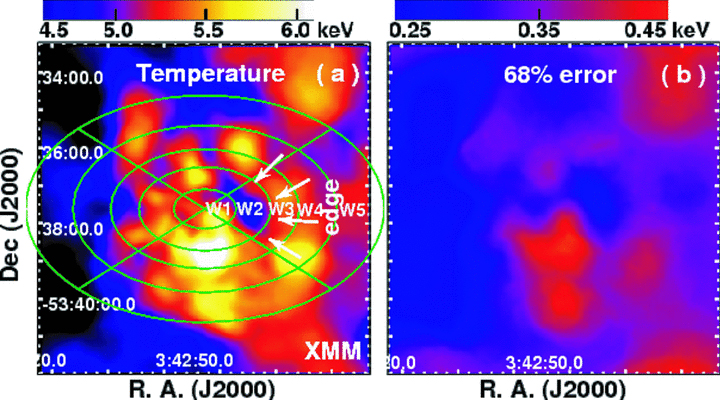}
    \caption{Temperature and error map from \cite{TempMap}.} 
    \label{fig:Wang}
 \end{figure}
\cite{TempMap} also produced a temperature map using the \textit{XMM-Newton} observation of the cluster, which is shown in Fig. \ref{fig:Wang}. A cool region is clearly detected west of this image. In order to compare this with the \textit{XMM-Newton} and eROSITA data, a second image of the temperature map was generated with the same dimensions and scale as \cite{TempMap}. This is shown in Fig. \ref{fig:TmapZ}. The location of the cool region is also detected in the eROSITA temperature map. Not only does the location of the cool clump agree, but the sharp increase in temperature west of this feature is also present in both maps. East of the cluster,  lower temperatures in the range $3.5 - 5.0$\,keV are found in both maps, as well as an increased temperature north and south of the central region. $5.5 - 6$\,keV gas is detected $\sim 6-7$\,arcmin in the north-west direction in both maps.
 \begin{figure}
     \centering
     \includegraphics[width=0.48\columnwidth]{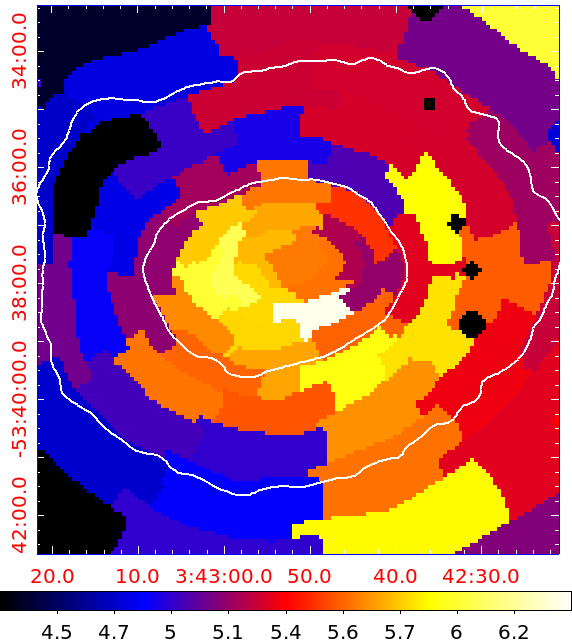}
     \includegraphics[width=0.48\columnwidth]{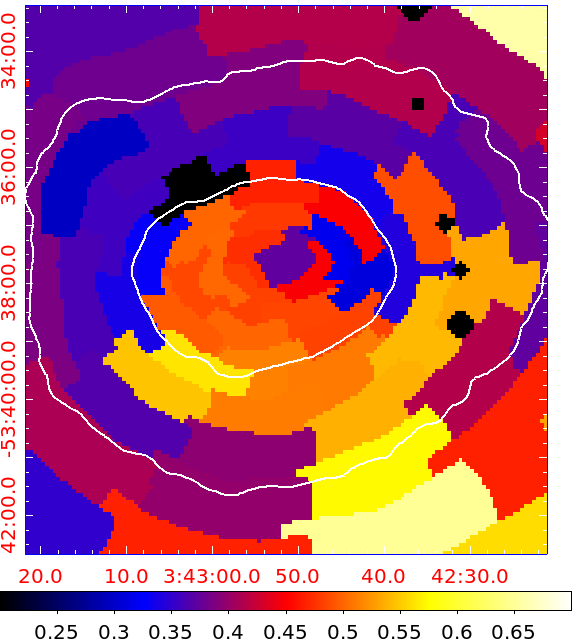}
     \caption{Temperature and error map in units of keV created using eROSITA data in the same coordinate dimensions as \cite{TempMap}. eROSITA A3158 contours are overlaid in white.} 
     \label{fig:TmapZ}
 \end{figure}
 
The lack of a cool core and the off-centre cool clump in both the \textit{XMM-Newton} and eROSITA temperature maps suggests that the Abell 3158 cluster is not in a relaxed state. Some merger activity may be ongoing in the cluster. The elongation of the overall X-ray surface brightness distribution in the east-west direction would be consistent with a merger-induced sloshing scenario in this direction.

Furthermore, in the region of the outer gas extension discussed in Section~\ref{morpho} and beyond it, we discover cool gas. The difference in temperatures on this large scale between this outer western part and the northern and southern parts of the cluster is marked. This supports the sloshing scenario; alternatively, the northern and southern region might be heated by outgoing transverse shocks from a previous merger.

\subsection{Velocity distribution}
The galaxy density maps clearly show that X-ray overdensities coincide with galaxy overdensities.  North of the field lies a population of galaxies that is classed as a probable separate substructure. This substructure corresponds to the extended source that is detected north of the cluster in both eROSITA and \textit{XMM-Newton}. The galaxies in this region are in the redshift range $0.070<z<0.077$.

South of field lies another galaxy overdensity that is classed as members of the A3158 cluster. This overdensity coincides with the location of both the SPT cluster and the southern extended source. This population of galaxies may be a substructure of A3158 that happens to be located at the same position as the SPT cluster.
It it possible that the Abell 3158 cluster is undergoing merger activity and that the substructure south of the FoV is being accreted by the cluster. 

\subsection{Mass determination}\label{section:Mass}
As discussed in \S \ref{Section:Spectral}, the mass of the cluster was estimated from the M-T relation \citep{Scaling}. The $M_{500}$ was estimated to be $4.03 \pm 0.49 \times 10^{14}$ $M_{\odot}$. This is slightly lower than the value determined by \cite{AngLiu} of $M_{500} = 4.53 \pm 0.11 \times 10^{14}$ $M_{\odot}$ which was determined using the same method, but the results are compatible within the errors. Furthermore, \cite{AngLiu} used \textit{Chandra} data to determine the mass from the M-T relation. \cite{Schellenberger15} showed that the temperature measurements from \textit{Chandra} are higher than those from \textit{XMM-Newton,} and as shown in this work, the eROSITA telescope temperature measurements are also lower than the \textit{Chandra} results. The observed offset is thus consistent with expectations. The hydrostatic mass was also determined in \cite{X-COPMass} and was determined to be $M_{500} = 4.26 \pm 0.18 \times 10^{14} M_{\odot}$. This is in between the masses determined in this work and by \citet{AngLiu}. Converting this into $M_{200}$, the mass estimated for A3158 from this work was $M_{200} = 6.20 \pm 0.75 \times 10^{14} M_{\odot}$. This agrees well with the hydrostatic mass value determined in \cite{X-COPMass} of $M_{200} = 6.63 \pm 0.39 \times 10^{14} M_{\odot}$.

In \S \ref{Section:VD} the mass of the cluster was calculated using the mass - velocity dispersion (M-$\sigma _{v}$) relation (summarised in Tab.~\ref{tab:Summary}), which resulted in systematically higher masses.
In dynamically active clusters, it is possible that the velocity dispersion is biased high. We have indicated that A3158 may be undergoing some merger activity, which might cause the velocity dispersion to be biased high. This would have the knock-on effect of causing the mass estimate to be biased high as well. 

Furthermore, it has been found that the observed temperature in merging clusters can be biased low relative to the total mass of the system because it takes time for the kinetic energy released during the merger to become completely thermalised \citep[e.g.][]{Kravtsov06}. This can influence the mass estimate from the M-T relation to have a low bias as well.
When these factors are taken into account, it is possible that the true M$_{200}$ of A3158 lies between $6.2 \times 10^{14} - 13.8 \times 10^{14} M_{\odot}$. 

\begin{table}[h]
    \begin{tabular}{c c c c}
    \hline\hline
    Publication & Method & Mass & Result ($10^{14} M_{\odot}$) \\
    \hline\hline
    This work & M-T & $M_{500}$ & $3.5 \pm 0.40$ \\
    Liu et al. & M-T & $M_{500}$ & $4.53 \pm 0.11$
    \\
    X-COP & Hydrostatic & $M_{500}$ & $4.26 \pm 0.18$ \\
    \hline
    This work & M-T & $M_{200}$ & $5.09 \pm 0.59$ \\
    X-COP & Hydrostatic & $M_{200}$ & $6.63 \pm 0.39$ \\
    This work & M-$\sigma_{v}$, Bocquet & $M_{200}$ & $13.8 \pm 2.5$ \\
    Moretti et al.* & M-$\sigma_{v}$, Bocquet & $M_{200}$ & $12.3 \pm 1.9$ \\
    This work & M-$\sigma_{v}$, Finn & $M_{200}$ & $19.8 \pm 2.3$\\
    Moretti et al. & M-$\sigma_{v}$, Finn & $M_{200}$ & $17.9$ \\

    \hline\hline
    \end{tabular}
    \caption{Summary of the masses determined in this work and the literature values with which they are compared. *Although \cite{OmegaWINGS} did not compute the mass using the \cite{Bocquet} relation, the $\sigma_{v}$ value they calculated was used with this relation in this work for the purpose of comparison. \cite{X-COPMass} is listed as X-COP.}
    \label{tab:Summary}
\end{table}

\section{Conclusions}\label{Section:Conclusion}
The galaxy cluster Abell 3158 was observed by the eROSITA observatory as a calibration source. We compared the 1D temperature, metal abundance, and normalisation profiles of the eROSITA observation and archival \textit{XMM-Newton} and \textit{Chandra} data. The temperatures measured with eROSITA, \textit{XMM-Newton,} and \textit{Chandra} agree at a level of $\lesssim 10\%$  and the profiles trend in the same direction, showing that the cluster lacks a cool core.
The metal abundance profile of the three telescopes shows a definitive decrease with increase in radius, as is expected, and they also agree at a level of  $\lesssim 10\%$ level. The eROSITA telescope provides tighter constraints on the metal abundance. The constraints also extend out to larger radii. The normalisation profile shows that the values obtained from the \textit{XMM-Newton} observation are higher than those from the eROSITA observation, but they are also within $\lesssim 10\%$. The normalisation values from \textit{Chandra}  agree well with the eROSITA data.

Many galaxies with spectroscopic redshifts lie in the Abell 3158 field. A redshift cut of $0.05<z<0.07$ was implemented to determine cluster members, and the velocity of these redshifts was calculated. The velocity dispersion of the member galaxies was determined, and a value for $M_{200,c}$ of $1.38 \pm 0.25 \times 10^{15} M_{\odot}$ was calculated using the mass - velocity dispersion relation. Using the M-T relation, we determined the $M_{200}$  to be $6.20 \pm 0.75 \times 10^{14} M_{\odot}$. Because dynamically active clusters typically have a velocity dispersion that is biased high and clusters undergoing merger activity host temperatures that are biased low, the true value of the cluster mass may lie between these values. The disagreement between the mass estimates further supports the claims that the cluster is undergoing merger activity. 

A population of galaxies with a spectroscopic redshift range $0.070<z<0.077$ was identified as a probable separate structure that corresponds to an extended source north of the main cluster. A similar extended source detected $1.85$ Mpc south of the main cluster also hosts an overdensity of galaxies. The galaxies located in the southern extended source are located at the same redshift as the main galaxy cluster. It is therefore likely being accreted onto the A3158 cluster. A high-redshift cluster previously discovered by SPT is also detected south of the cluster centre.

The 2D temperature map of the cluster showed that the cluster does not have a cool core. A cool clump lies west of the central region; this was observed before. This matches the temperature map presented in \cite{TempMap} closely. The off-centre cool clump and the lack of a cool core suggest that the cluster is not relaxed and may be undergoing some merger activity. The bow-shaped edge located near the cool gas clump $\sim 137$ kpc west of the cluster centre we detected is also consistent with the previously discovered cold front in \textit{XMM-Newton} and \textit{Chandra} observations. The extension of gas $\sim 685$ kpc west of the cluster centre is a new discovery and supports the idea that the cluster is not relaxed, but is undergoing merger activity. The surface brightness and temperature map analyses together clearly confirm that this is a disturbed cluster.

\begin{acknowledgements}
We would like to thank the anonymous referee for very useful comments that improved the manuscript. This work is based on data from eROSITA, the soft X-ray instrument aboard SRG, a joint Russian-German science mission supported by the Russian Space Agency (Roskosmos), in the interests of the Russian Academy of Sciences represented by its Space Research Institute (IKI), and the Deutsches Zentrum für Luft- und Raumfahrt (DLR). The SRG spacecraft was built by Lavochkin Association (NPOL) and its subcontractors, and is operated by NPOL with support from the Max Planck Institute for Extraterrestrial Physics (MPE).

The development and construction of the eROSITA X-ray instrument was led by MPE, with contributions from the Dr. Karl Remeis Observatory Bamberg \& ECAP (FAU Erlangen-Nuernberg), the University of Hamburg Observatory, the Leibniz Institute for Astrophysics Potsdam (AIP), and the Institute for Astronomy and Astrophysics of the University of Tübingen, with the support of DLR and the Max Planck Society. The Argelander Institute for Astronomy of the University of Bonn and the Ludwig Maximilians Universität Munich also participated in the science preparation for eROSITA.

The eROSITA data shown here were processed using the eSASS software system developed by the German eROSITA consortium.

Part of this work has been funded by the Deutsche Forschungsgemeinschaft (DFG, German Research Foundation) – 450861021.

This work was supported in part by the Fund for the Promotion of Joint International Research, JSPS KAKENHI Grant Number 16KK0101.

This research has made use of the NASA/IPAC Extragalactic Database (NED), which is funded by the National Aeronautics and Space Administration and operated by the California Institute of Technology.

This research has made use of the VizieR catalogue access tool, CDS, Strasbourg, France (DOI: 10.26093/cds/vizier). The original description of the VizieR service was published in A\&AS 143, 23

We would like to express thanks to Professor Haiguang Xu for allowing us to show the temperature map created using \textit{XMM-Newton} published in \citealt{TempMap}.
\end{acknowledgements}


\bibliographystyle{aa} 
\bibliography{bib}

\begin{appendix}
\label{sec:A1}
\section{Light curves}
\begin{figure}
    \centering
    \includegraphics[width=\columnwidth]{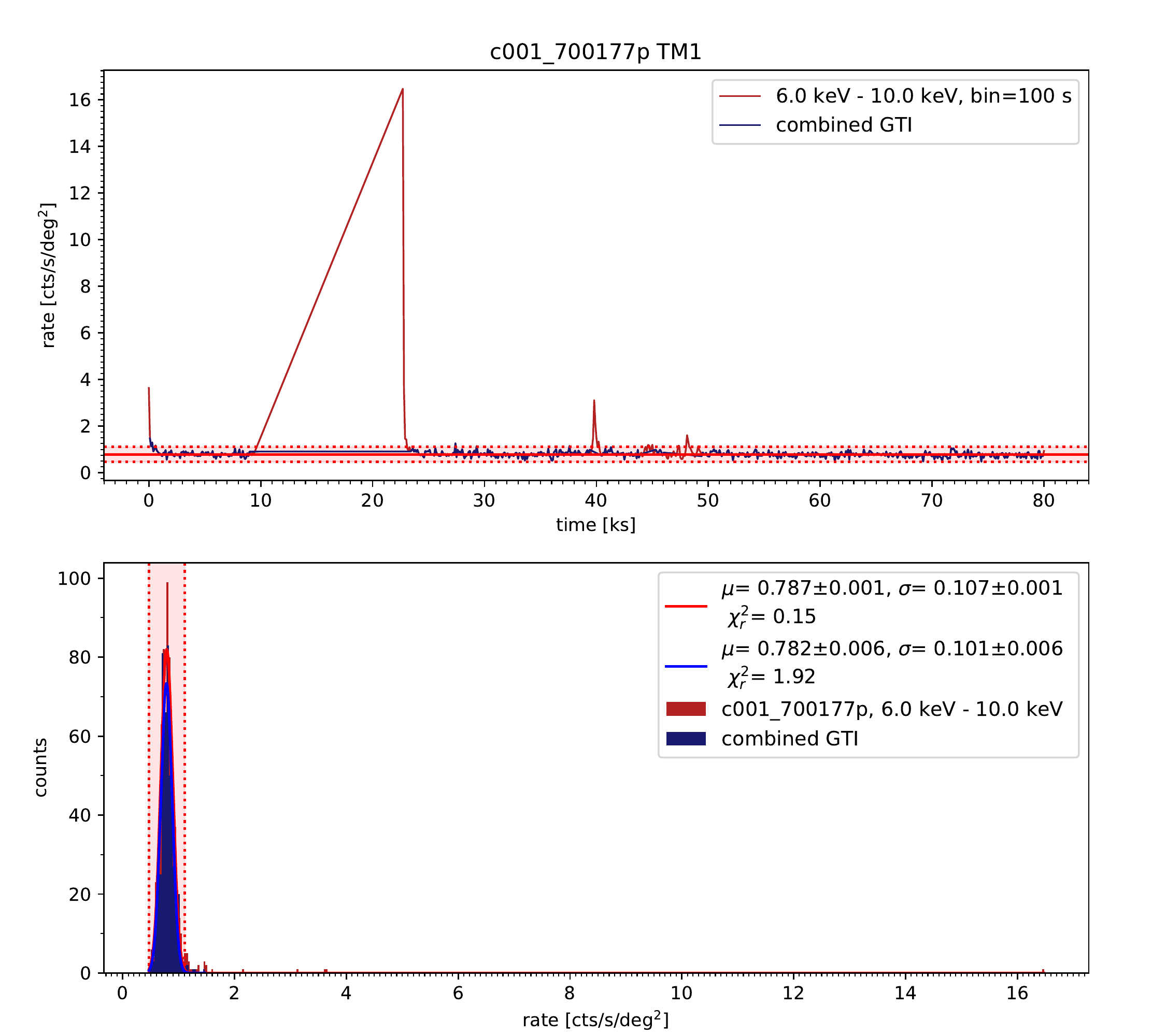}
    \includegraphics[width=\columnwidth]{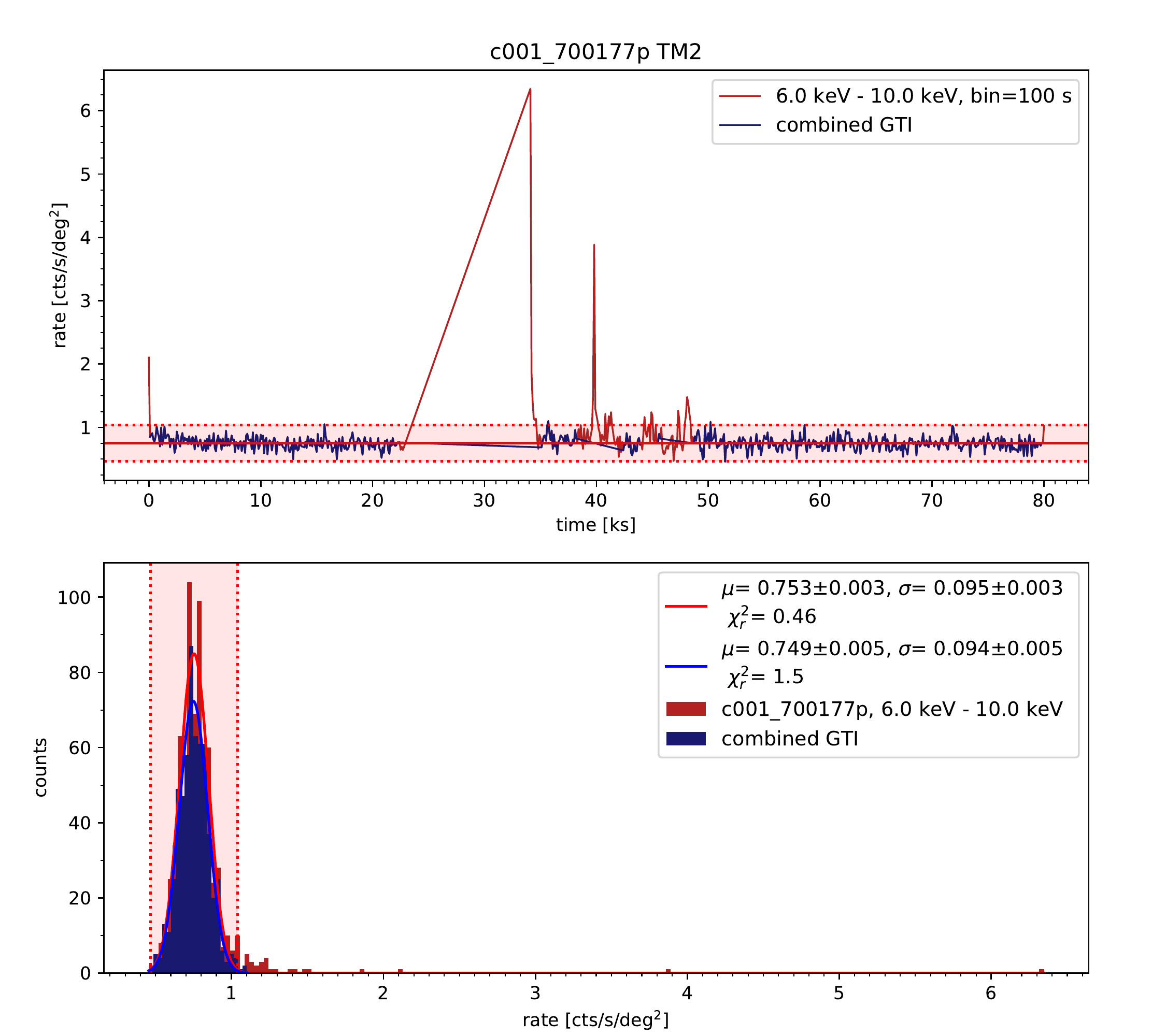}
    \includegraphics[width=\columnwidth]{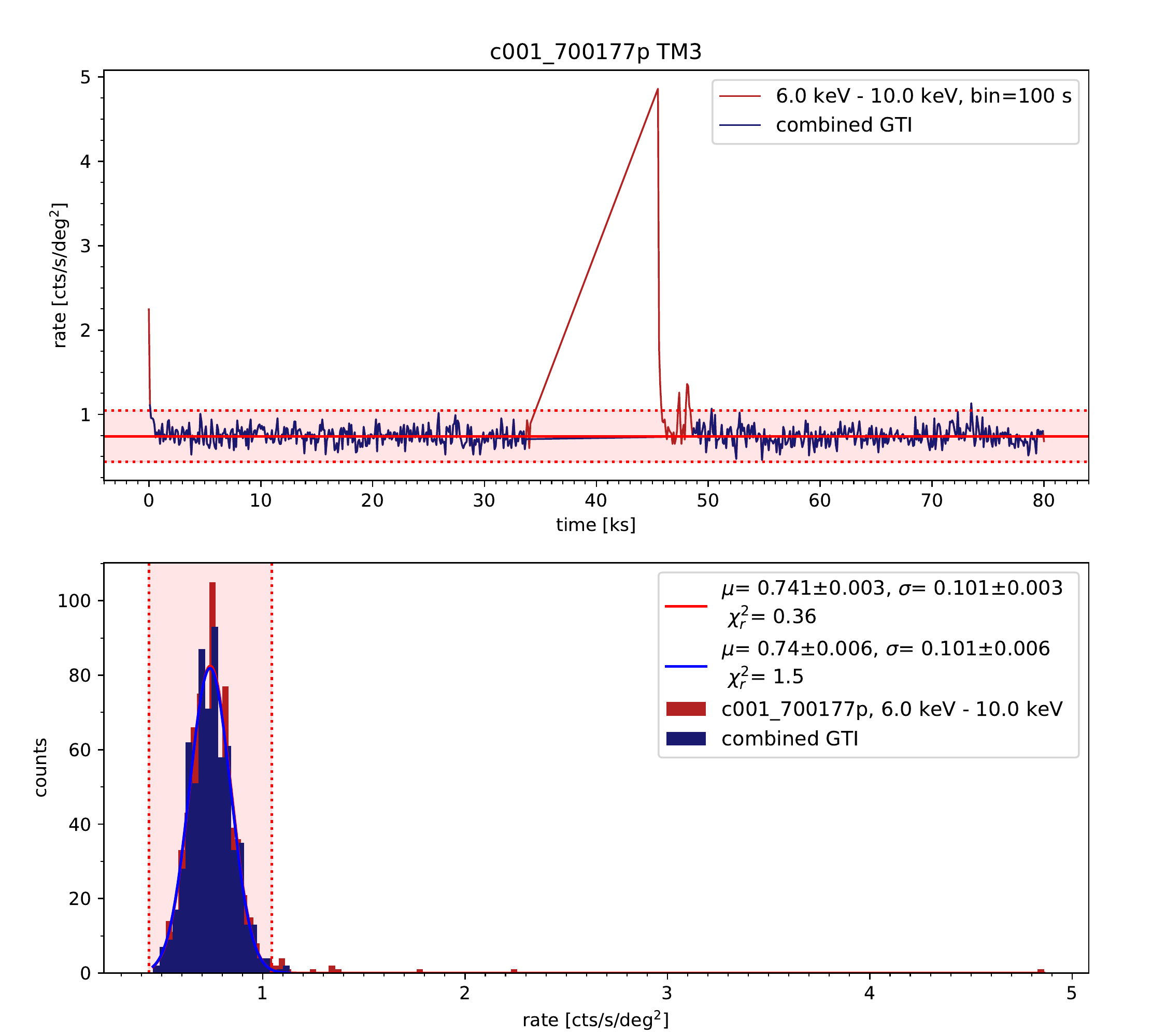}
    \caption{Light  curves of TMs 1, 2, and 3.}
    \label{fig:A1}
\end{figure}

\begin{figure}
    \centering
    \includegraphics[width=\columnwidth]{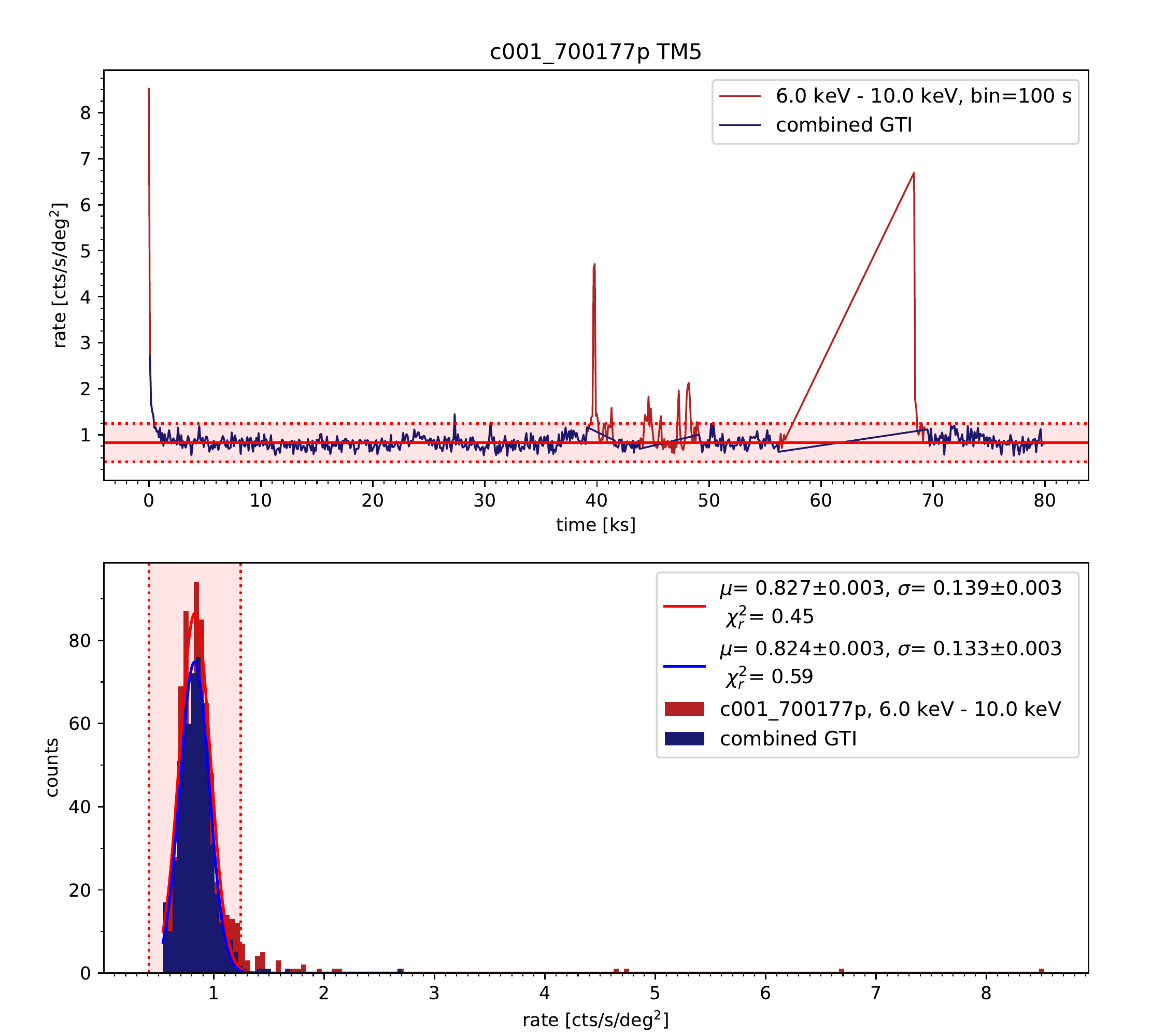}
    \includegraphics[width=\columnwidth]{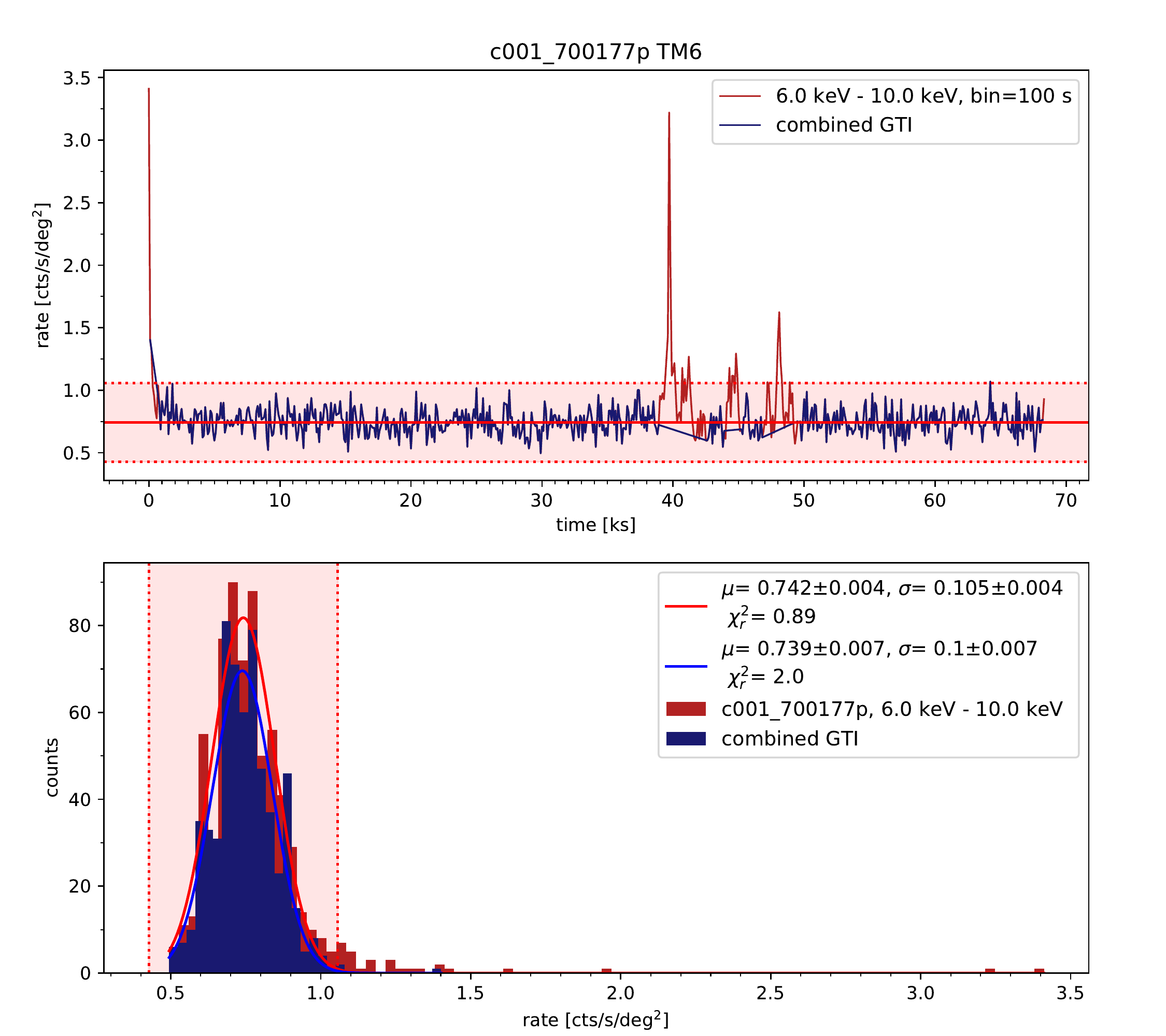}
    \includegraphics[width=\columnwidth]{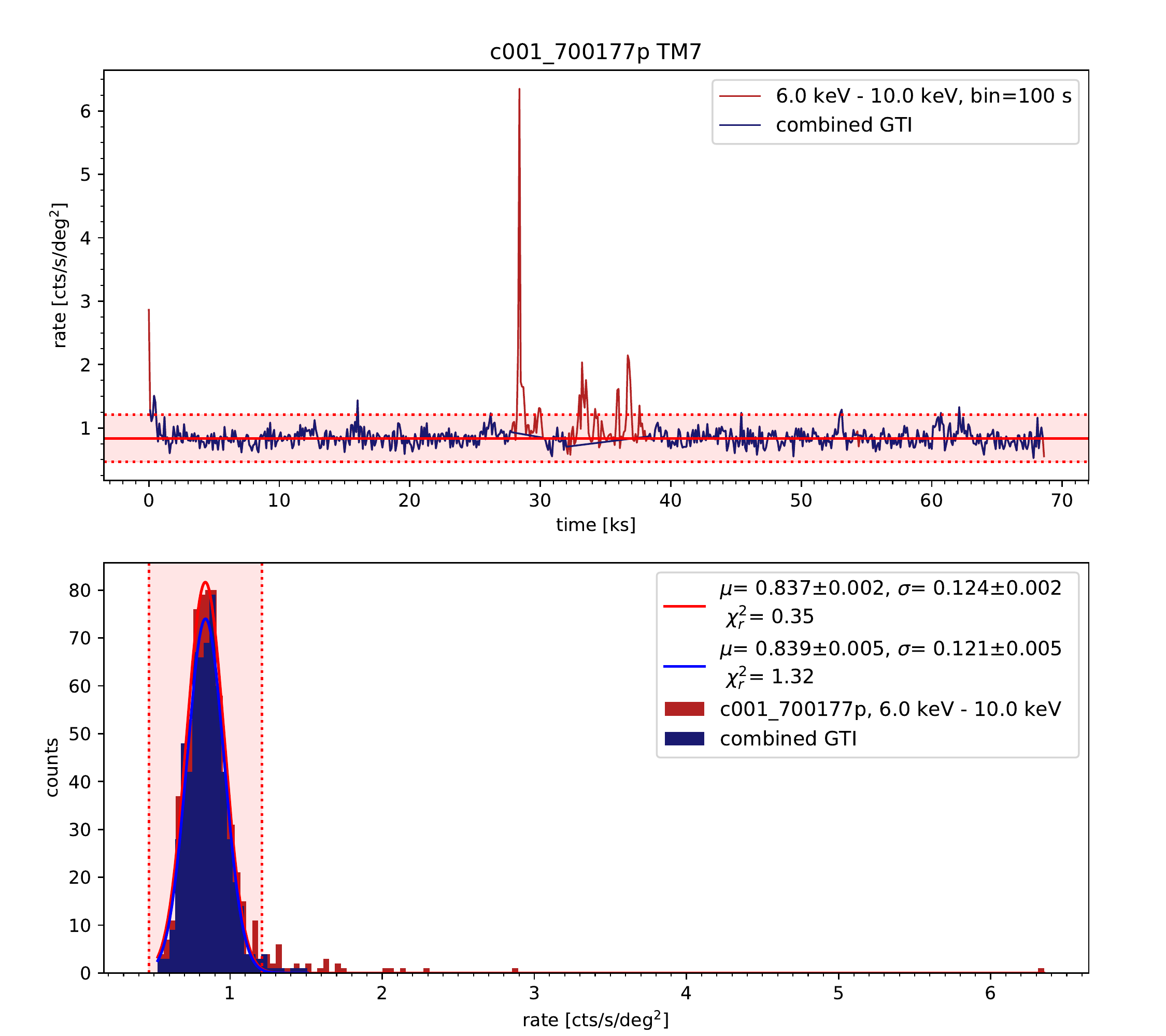}
    \caption{Light curves of TMs 5, 6, and 7}
    \label{fig:A2}
\end{figure}

Fig. \ref{fig:A1} and \ref{fig:A2} show the light curves of the remaining six telescopes with a binning of $100$\,s in the $6-9$\,keV energy range. The flare between $37-43$\,ks is visible in most of the light curves, as are the excluded time periods during which the FWC data were observed for each telescope. 

\section{Ratio plot}
\begin{figure}
    \centering
    \includegraphics[width=\columnwidth]{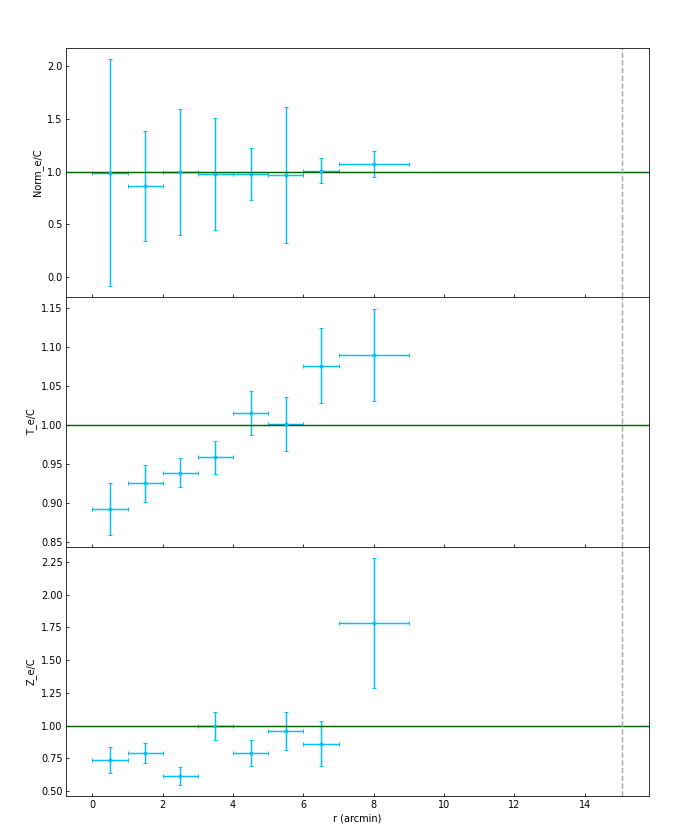}
    \caption{Plot of the normalisation, temperature, and abundance ratios of eROSITA/\textit{Chandra}. The dashed vertical grey line is the $r_{500}$ of the cluster. }
    \label{fig:A3}
\end{figure}

Fig. \ref{fig:A3} shows the ratio plot of the profiles comparing eROSITA and \textit{Chandra} results. 
\end{appendix}
\end{document}